\documentclass[11pt,fleqn]{article}
\usepackage{amssymb}

\usepackage{graphicx}
\usepackage{amsmath}
\usepackage{times}
\usepackage{hyperref}

\newtheorem{theorem}{Theorem}

\newtheorem{remark}[theorem]{Remark}

\input{tcilatex}
\begin{document}

\title{A method to compute periodic sums}
\author{Nail A. Gumerov\thanks{%
Corresponding Author. Also at Center for Micro and Nanoscale Dynamics of
Dispersed Systems, Bashkir State University, Ufa, Russia; and at Fantalgo,
LLC., Elkridge, MD 21075; E-mail: \texttt{gumerov@umiacs.umd.edu}; Phone:
+1-301-405-8210; Fax: +1-301-314-9658; web: \texttt{%
http://www.umiacs.umd.edu/users/gumerov}} \ \ and \ Ramani Duraiswami\thanks{%
Also Department of Computer Science, and at Fantalgo, LLC. E-mail:\texttt{%
ramani@umiacs.umd.edu}, web: \texttt{http://www.umiacs.umd.edu/users/ramani}}
\\
Institute for Advanced Computer Studies\\
University of Maryland, College Park}
\date{October 15, 2013 \\
revised \ March 30, 2014}
\maketitle

\begin{abstract}
In a number of problems in computational physics, a finite sum of kernel functions centered at $N$ particle locations located in a box in three dimensions must be extended by imposing periodic boundary conditions on box boundaries. Even though the finite sum can be efficiently computed via fast summation algorithms, such as the fast multipole method (FMM), the periodized extension is usually treated via a different algorithm, Ewald summation, accelerated via the fast Fourier transform (FFT). A different approach to compute this periodized sum just using a blackbox finite fast summation algorithm is presented in this paper. The method splits the
periodized sum in to two parts. The first, comprising the contribution of all points outside a large sphere enclosing the box, and some of its neighbors, is approximated inside the box by a collection of kernel functions (``sources'') placed on the surface of the sphere or using an expansion in terms of spectrally convergent local basis functions. The
second part, comprising the part inside the sphere, and including the box
and its immediate neighborhood, is treated via available summation
algorithms. The coefficients of the sources are determined by least squares
collocation of the periodicity condition of the total potential, imposed on
a circumspherical surface for the box. While the method is presented in
general, details are worked out for the case of evaluating electrostatic
potentials and forces. Results show that when used with the FMM, the
periodized sum can be computed to any specified accuracy, at an additional cost of the order of the  free-space FMM. Several technical
details and efficient algorithms for auxiliary computations are provided, as
are numerical comparisons.
\end{abstract}

\subsubsection*{Keywords}

periodic sums; fast multipole method; Ewald summation; GPU computing; kernel
independent methods; molecular dynamics; long-range interactions

\subsubsection*{Acknowledgments}

Work partially supported by the following sources: AFOSR under MURI Grant
W911NF0410176 (PI Prof.~J.~G. Leishman, monitor Dr.~D. Smith); by NSF award
1250187 (PI: Prof.~B.~Balachandran); by Grant G34.31.0040 (PI:
Prof.~I.~Akhatov) of the Russian Ministry of Education \& Science.; and by
Fantalgo, LLC.

\newpage

\section{Introduction}

Many problems in physics, chemistry and materials science lead to a
free-space finite ``particle'' sum of $N$ functions, $K$, centered at
locations $\mathbf{x}_i$ $\in \Omega _{0}\subset \mathbb{R}^{3}$, where $%
\Omega _{0}$ is a rectangular box $d_{1}\times d_{2}\times d_{3}$ centered
at the origin of the reference frame 
\begin{equation}
\tilde{\phi} \left( \mathbf{y}\right) = \sum_{i=1}^{N}q_{i}K\left(\mathbf{y}%
- \mathbf{x}_{i}\right).  \label{r2}
\end{equation}
For evaluation at $N$ locations $\mathbf{y}$, this sum has a quadratic cost.
There are efficient and arbitrarily accurate approximation algorithms for
this summation (e.g., the fast multipole method, FMM \cite{Greengard1987}).

Often, an extension to this sum for $\tilde{\phi}$ must be computed in which
periodic boundary conditions are enforced on box boundaries, resulting in
the potential $\phi $. This can be evaluated by replacing the sum (\ref{r2})
with the infinite sum 
\begin{equation}
\phi \left( \mathbf{y}\right) =\sum_{\mathbf{p}}\sum_{i=1}^{N}q_{i}K\left( 
\mathbf{y}-\mathbf{x}_{i}+\mathbf{p}\right) ,\quad \mathbf{p}\in \mathbb{P}%
=\left\{ \left( i_{1}d_{1},i_{2}d_{2},i_{3}d_{3}\right) :\left(
i_{1},i_{2},i_{3}\right) \in \mathbb{Z}^{3}\right\} .  \label{2}
\end{equation}
For some functions $K$, such as those representing the field of an
electrostatic charge, this infinite sum may be divergent or conditionally
convergent. In this case certain side conditions may be needed to compute a
physically relevant sum. Usually such infinite sums are performed using
Fourier-transform based Ewald summation \cite{Ewald:AnnPhys1921}, which is
accelerated via the FFT. This method is described briefly in Appendix D.
Accounting for all pairwise interactions the method can achieve $O(N\log N)$
complexity, for $N$ particles in the box $\Omega _{0}$ which is periodically
replicated over the full space \cite{Darden:JChP1993, Essmann:JChP1995}.
Because of the technique used for grid-to-particle interpolation these
methods are usually low-order. A high-order accurate Gaussian interpolation
based Ewald summation algorithm was recently presented in \cite
{Lindbo:JCP2010, Lindbo:JCP2011}.

Another scalable algorithm, which can be employed for computation of
periodic sums (\ref{2}) is the FMM. A criticism of FMM algorithms has been
that they are relatively harder to implement, combining the need for
efficient data structures, careful analysis and computation of special
functions, and mixed memory access patterns. Nevertheless, several
open-source and commercial packages implementing the FMM for standard
kernels in free space have become available. The FMM is used less often in
practice to compute periodic sums, even though several methods to handle
periodic boundary conditions using extensions to the basic FMM have been
proposed, starting from the first publication of the algorithm \cite
{Greengard1987, Schmidt:JSP1991, Christiansen:JCP1993, Hamilton1995,
Lambert:JCP1996, Figueirido:JChemPhys1997, Amisaki:JCC2000, Otani:IJMCE2006,
Otani:JCP2008, Barnett:JCP2010, Langston:CAMCOS2011}. 
Some authors (e.g. \cite{Amisaki:JCC2000}) claim that the overhead in the FMM for
computation of periodic sums can be negligibly small. 
This is true under a
few conditions, but not in others. This is discussed in detail in a section entitled Discussion.
Analysis of the FMM and its plane-wave
variant, which is better-suited tor large $N$ and parallel architectures
than the smooth particle mesh Ewald algorithm, is presented in \cite
{Kia:JCP2008}. However, all these methods require constructing a new and
different algorithm -- a periodic variant, for which optimized
implementations are not in general available. Furthermore, some of these algorithms may not extend easily
to non-cubic domains.


It can be mentioned in this context that any FMM has a so-called break-even
point, $N_{b}$, such that at $N<N_{b}$ brute force summation (\ref{r2}) is
faster than the FMM computation. The value of $N_{b}$ depends on the method used, implementation, required accuracy (the FMM is an approximate
algorithm), and hardware. It may vary in a range $N_{b}\sim 10^{2}-10^{4}$,
which is of the order of $N$ for some practical problems (e.g. for some
molecular dynamics simulations). In such cases, the use of periodic FMM just
for the purpose of computing of infinite sums (\ref{2}) even if the overhead
for periodization will be zero is questionable if some other periodization
method, which can use finite brute force summation algorithms is available.

Special purpose hardware such as graphics processors or heterogeneous
CPU/GPU architectures also allow the fast computation of finite sums, either
via brute force summation \cite{Stone:JCC2007}, or via the mapping of the
FMM onto these architectures \cite{Gumerov:JCP2008, Hamada:SC2009,
Hu:SC2011, Yokota:CPC2013}. Yokota et al. \cite{Yokota:CPC2013} favorably
compare a large scale FMM-based vortex element computations with a direct
numerical simulation via periodic pseudospectral methods. Their simulations
could have been faster and more accurate -- the FMM was executed on a finite
system composed of 3$^{3}$ images, which while not being truly periodic also
makes using the FMM significantly more expensive.

The problem this paper seeks to address is: \emph{Given a black-box fast
summation algorithm (FSA) for computing finite sums with a
given kernel} $K\left( \mathbf{y}-\mathbf{x}_{i}\right) $ that is available to the user, \emph{is it
possible to compute the same sum with periodic boundary conditions without
any modification of the FSA?} We provide a positive answer to this question.
Our algorithm has the same cost as the FSA, and can be computed to any user
specified accuracy $\epsilon $, and does not use the FFT. The basic idea of
the method is to divide the sum (\ref{2}) in to two parts. One part computes
a finite sum of particles that lie within a sphere centered at the box. This
is computed using the available FSA. The other part, is an approximation of
the field within the box due to all particles outside the sphere. The field
due to these sources can be represented within the box in terms of local
expansions. Such local expansions have also been proposed in other attempts
to extend the FMM to periodic systems, but are there derived by explicit
translation of multipole expansions from outside the box of interest into
it. In our method, we propose to determine the coefficients directly from
the periodicity conditions on the potential, which results in solution of a
relatively small overdetermined function-fitting problem, easily solved via
standard algorithms -- e.g., rank-revealing $QR$ decomposition. This step is
in the spirit of the ``kernel-independent'' FMM methods \cite{Ying:JCP2004,
Ying:JCP2006}. In this context we should mention Ref. \cite{Barnett:JCP2010}%
. Even though it is dedicated to the Helmholtz equation in two dimensions, a
method for ``periodization'' of free-space solutions similar in spirit to that presented here was proposed and tested. Moreover, the cited paper contains an additional  ``periodization'' method based on boundary integrals,
which can be tried for different kernels and space dimensionality (also see 
\cite{Barnett:BIT2011}, where periodization along one dimension is performed).

We present this ``periodization'' approach in a general setting, but focus
computational examples on the evaluation of the electrostatic potential $%
\phi $ and its gradient $\nabla \phi $ at $M$ evaluation points $\mathbf{y}%
_{j}$ $\in \Omega_{0}\subset \mathbb{R}^{3}$ due to $N$ charged particles of
charge $q_{i}$ placed in $\Omega_{0}$, and subject to periodic boundary
conditions on $\partial \Omega_{0}$. This reduces to computing the infinite,
conditionally convergent, sum (\ref{r2}), with kernel function $K$: 
\begin{equation}
K\left( \mathbf{y}-\mathbf{x}\right) =\frac{1}{\left| \mathbf{y}-\mathbf{x}%
\right| },\quad \mathbf{y}\neq \mathbf{x;\quad }K\left( \mathbf{y}-\mathbf{x}%
\right) =0,\quad \mathbf{y}=\mathbf{x}, \qquad \sum_{i=1}^{N}q_{i}=0.
\label{1}
\end{equation}
and the net charge in each box being zero. Of course, in practical
applications, such as in molecular dynamics, there will be other
computations beyond the sum necessary  to stabilize the overall
computations. This paper does not consider these, focusing on the
electrostatic sum at a single time step.

The periodization method could be easily applied to other kernels for which
a ``fast summation algorithm'' (FSA) is available. As long as the periodic
sum makes sense, and the kernel $K$ can be expanded over some local basis,
the proposed method should work. Conditions similar to the charge neutrality in (\ref
{1}) may be necessary. The proposed method can also be applied for 2D
problems, though we present it in 3D. Also, the periodic extension of the
computational domain (box) may be performed only along one or two
coordinates, for which Ewald summation may have problems.

\section{Proposed method}
\subsection{Periodization}
\begin{figure}[tbh]
\begin{center}
\includegraphics[trim=0.2in 0.5in 2.5in 0.2in,
width=4.5in]
{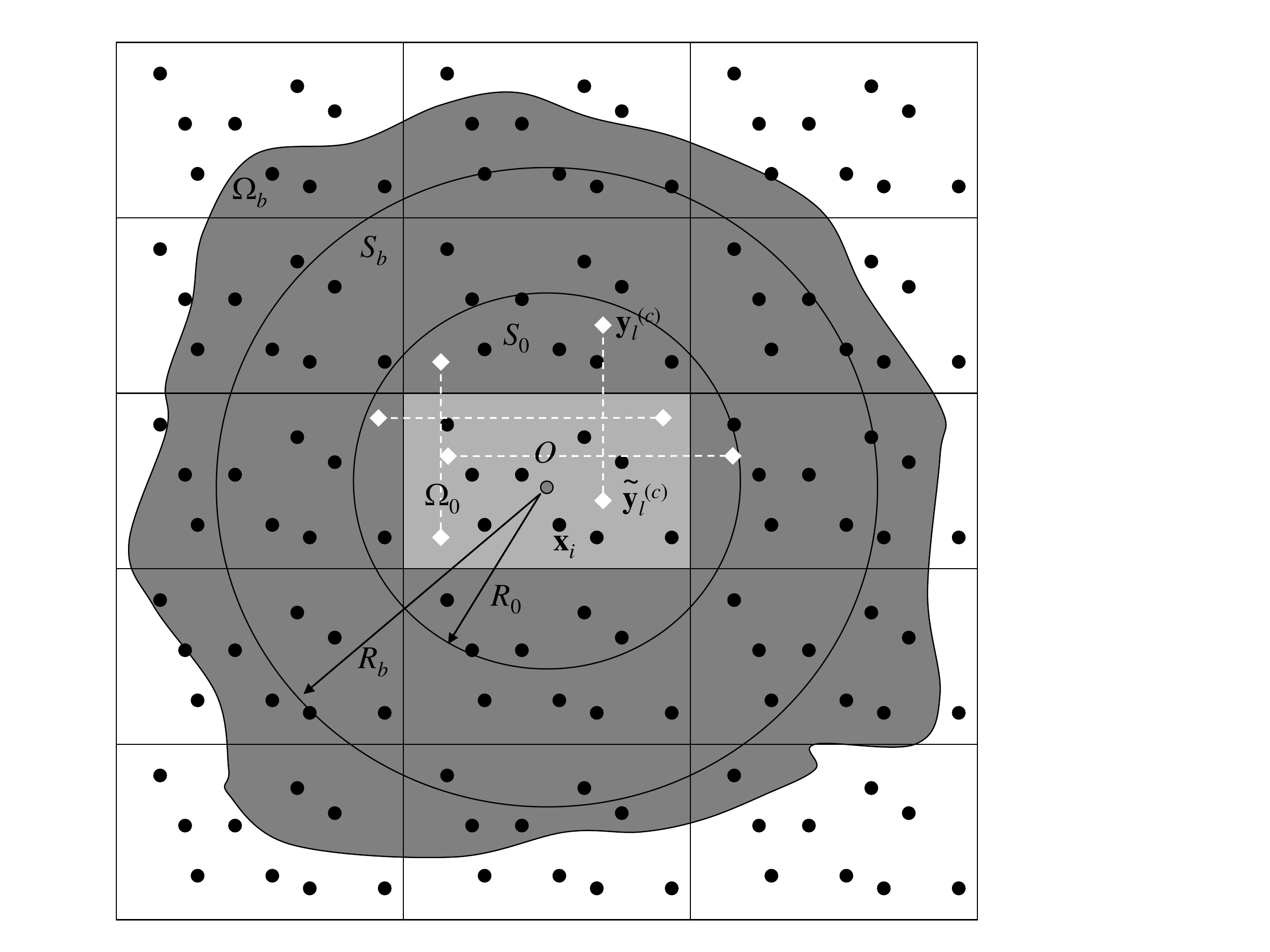}
\end{center}
\caption{The particle sum (\ref{2}) in the box $\Omega_0$, subject to
periodic boundary conditions on the box boundary, is to be computed to a
specified accuracy $\protect\epsilon$. The boundary condition can be
enforced by the image method, which results in an infinite set of copies of
the box $\Omega_0$. The proposed method divides the sum into a near-field
component in $\Omega_b$ and a far-field sum.}
\label{Fig1}
\end{figure}

The box on which the periodic sum is to be computed is denoted $\Omega _{0}$%
. Let $S_0$ be a ball of radius $R_0$ centered at the center of the box $%
\Omega _{0}$ and containing it. Let $S_{b}$ be another ball with the same
center and radius $R_{b}>R_{0}$ (see Fig. \ref{Fig1}). We denote as $%
\Omega_b $ a finite region which includes the ball $S_b$ (the minimal $%
\Omega_b$ is the ball $S_b$). We decompose the infinite sum as 
\begin{equation}
\phi \left( \mathbf{y}\right) =\phi _{near}\left( \mathbf{y}\right) +\phi
_{far}\left( \mathbf{y}\right) ,\quad \phi _{near}\left( \mathbf{y}\right)
=\sum_{\mathbf{x}_{j}\in \Omega _{b}}q_{j}K\left( \mathbf{y}-\mathbf{x}%
_{j}\right) ,\quad \phi _{far}\left( \mathbf{y}\right) =\sum_{\mathbf{x}%
_{j}\notin \Omega _{b}}q_{j}K\left( \mathbf{y}-\mathbf{x}_{j}\right) ,
\label{3}
\end{equation}
where $\phi _{near}\left( \mathbf{y}\right) $ is to be computed using the
FSA (assumed to be the FMM in the sequel) to the specified accuracy $%
\epsilon $ while $\phi _{far}\left( \mathbf{y}\right) $ is computed by some
other method at least to the same accuracy. The sources in the infinite
domain are indexed as $\mathbf{x}_{j}=\mathbf{x}_{i}-\mathbf{p}$, $%
q_{j}=q_{i}$ for appropriate vectors $\mathbf{p\in }$ $\mathbb{P}$ (see Eq. (%
\ref{2})).

To apply the FMM to the kernel function $K\left( \mathbf{y}\right) $ it
should be possible to approximate it via a convergent series over a set of
local basis functions $\left\{ R_{t}\left( \mathbf{y}\right) \right\} $.
This means that for any source point $\mathbf{x}_{j}\notin \Omega _{b}$ we
have the factorization 
\begin{equation}
K\left( \mathbf{y}-\mathbf{x}_{j}\right) =\sum_{t=1}^{P}B_{t}\left( \mathbf{x%
}_{j}\right) R_{t}\left( \mathbf{y}\right) +\epsilon _{j}^{(P)},\quad \left| 
\mathbf{x}_{j}\right| >R_{b},\quad \left| \mathbf{y}\right| <R_{0},
\label{4}
\end{equation}
where $P$ is the number of terms retained in the infinite series, $%
B_{t}\left( \mathbf{x}_{j}\right) $ are the expansion coefficients, and $%
\epsilon _{j}^{(P)}$ is the truncation error depending on $\left| \mathbf{x}%
_{j}\right| $ and $R_{0}$. The basis functions can be some standard choices,
or, as in the kernel independent FMM \cite{Ying:JCP2006}, can be taken to be
a collection of kernel functions, centered outside the region of
approximation 
\begin{equation}
R_{t}\left( \mathbf{y}\right) =K\left( \mathbf{y}-\mathbf{x}%
_{t}^{(s)}\right) ,\quad \left| \mathbf{x}_{t}^{(s)}\right| >R_{b},
\label{5}
\end{equation}
where $\mathbf{x}_{t}^{(s)}$ are a collection of sources located outside
ball $S_{b}$. This method has the flavor of ``equivalent-source'' methods.
Since the function $K$ depends only on the distance between its argument, it
is a ``radial basis function'', or RBF \cite{Buhmann:Book2003}. While any
approximation scheme may be used, the use of kernel $K$ as RBF is dictated
by the fact that this kernel satisfies the underlying equation (e.g. the
Laplace equation), so the approximation to the field via the sum of such
RBFs also satisfies the equation (if it is linear and space invariant).
Substituting Eq. (\ref{4}) into the expression for $\phi _{far}\left( 
\mathbf{y}\right) $ from Eq. (\ref{3}), we obtain 
\begin{eqnarray}
\phi _{far}\left( \mathbf{y}\right)  &=&\sum_{t=1}^{P}C_{t}R_{t}\left( 
\mathbf{y}\right) +\epsilon ^{(P)},\quad   \label{6} \\
C_{t} &=&\sum_{\mathbf{x}_{j}\notin \Omega _{b}}q_{j}B_{t}\left( \mathbf{x}%
_{j}\right) ,\quad \epsilon ^{(P)}=\sum_{\mathbf{x}_{j}\notin \Omega
_{b}}q_{j}\epsilon _{j}^{(P)}.  \notag
\end{eqnarray}
A necessary condition for our method is convergence of both infinite sums $%
C_{t}$ and $\epsilon ^{(P)}$. The problem of computation of $\phi
_{far}\left( \mathbf{y}\right) $ has been reduced to that of determination
of $P$ fitting coefficients $C_{t}$. These can be determined via
least-squares collocation as follows. Consider a set of $L>P$ check points, $%
\mathbb{Y}^{(c)}\subset S_{0}\backslash \Omega _{0}$. A point $\mathbf{y}%
_{l}^{(c)}\in \mathbb{Y}^{(c)}$ has two properties:first,$\left| \mathbf{y}%
_{l}^{(c)}\right| <R_{0}$, and, second, that there exists $\mathbf{p\in }%
\mathbb{P}$ such that point $\widetilde{\mathbf{y}}_{l}^{(c)}=\mathbf{y}%
_{l}^{(c)}+\mathbf{p}\in S_{0}$ (see Fig. \ref{Fig1}). This means that 
\begin{equation}
\phi \left( \mathbf{y}_{l}^{(c)}\right) =\phi \left( \widetilde{\mathbf{y}}%
_{l}^{(c)}\right) ,\quad l=1,...,L.  \label{7}
\end{equation}
In terms of decomposition (\ref{3}) and representation of the far field (\ref
{6}) this system can be rewritten as 
\begin{eqnarray}
\sum_{t=1}^{P}A_{lt}C_{t} &=&f_{l}+\epsilon _{l}^{(P)},\quad l=1,...,L,
\label{8} \\
A_{lt} &=&R_{t}\left( \mathbf{y}_{l}^{(c)}\right) -R_{t}\left( \widetilde{%
\mathbf{y}}_{l}^{(c)}\right) ,\quad f_{l}=\phi _{near}\left( \widetilde{%
\mathbf{y}}_{l}^{(c)}\right) -\phi _{near}\left( \mathbf{y}_{l}^{(c)}\right) 
\notag
\end{eqnarray}
where $\left| \epsilon _{l}^{(P)}\right| =\left| \epsilon ^{(P)}\left( 
\widetilde{\mathbf{y}}_{l}^{(c)}\right) -\epsilon ^{(P)}\left( \mathbf{y}%
_{l}^{(c)}\right) \right| \leqslant 2\max_{\mathbf{y\in }S_{0}}\left|
\epsilon ^{(P)}\left( \mathbf{y}\right) \right| .$ We have $L$ linear
equations in $P$ unknowns $C_{1},...,C_{P}$. As we are not constrained with
the size of the set of the set points, $L>P$ can be selected to provide a
substantial oversampling, so minimization of functional 
\begin{equation}
F\left( C_{1},...,C_{P}\right) =\sum_{l=1}^{L}\left(
\sum_{t=1}^{P}A_{lt}C_{t}-f_{l}\right) ^{2},  \label{9}
\end{equation}
should take care about the ``noise'' introduced into the approximation due
to $\epsilon _{l}^{(P)}$. The least square minimization procedure is well
known and formally it results in solution 
\begin{equation}
\mathbf{C=A}^{\dagger }\mathbf{f,\quad A}^{\dagger }=\left( \mathbf{A}^{T}%
\mathbf{A}\right) ^{-1}\mathbf{A}^{T},  \label{10}
\end{equation}
where $\mathbf{C=}\left\{ C_{t}\right\} $ and $\mathbf{f=}\left\{
f_{l}\right\} $ are organized as column vectors of size $P$ and $L$,
respectively and $\mathbf{A}^{\dagger }$ is the $P\times L$ matrix, which is
the pseudoinverse of $\mathbf{A}$, and superscript $T$ denotes
transposition. Note that this notation is formal, and is not the way the
least-squares problem is solved in practice. Rather a stable algorithm such
as the rank-revealing QR decomposition \cite{Golub:Book1996} is used, while
it is strongly recommended to precompute and store matrix decompositions in the ``set'' part of the algorithm to reduce the cost of the ``get'' part.

The known coefficients $\mathbf{C}$ allow computation of $\phi _{far}\left( 
\mathbf{y}\right) $ and can be added to the $\phi _{near}$ obtained via the
FSA. However, some technical details need to be specified. In the next
section we provide analysis and details for the important case of the
Coulombic kernel (\ref{2}).

A similar collocation of kernel based RBF expansions at a relatively small
amount of the check points is also used in the ``kernel independent'' FMM 
\cite{Ying:JCP2004} with basis functions (\ref{5}). There, the collocation
is at the level of the boxes in the FMM octree data structure, and fitting
takes the place of expansions and translations. Here, we collocate the
differences of the overall solution at a set of check points $\mathbf{y}%
_{l}^{(c)}$ and at their periodic images $\widetilde{\mathbf{y}}_{l}^{(c)}$,
at the level of the overall domain to determine the expansion coefficients
for $\phi_{far}$.

\subsection{Check point set}

\begin{figure}[tbh]
\begin{center}
\includegraphics[trim=0.2in 3.75in 0.2in 0.25in, width=6.0in]
{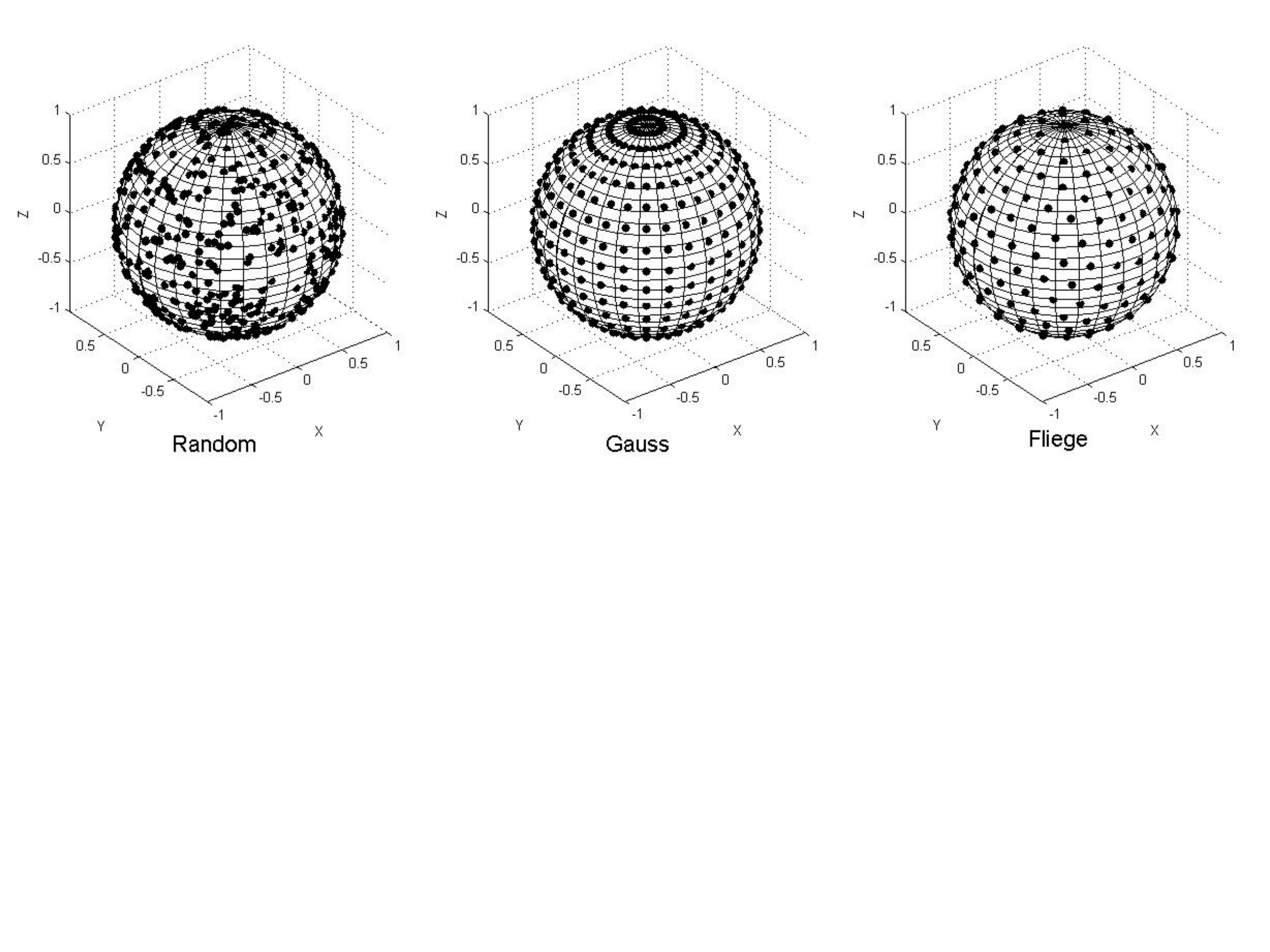}
\end{center}
\caption{The check point distributions over the surface of a unit sphere for 
$p=16$: random distribution ($L=3p^{2}$ points), the Gauss spherical grid ($%
L=2p^{2}-p$), and the Thomson points ($L=p^{2}$).}
\label{CheckPoints}
\end{figure}
Selection of an optimal set of check points $\mathbb{Y}^{(c)}\subset
S_{0}\backslash \Omega _{0}$ is not trivial. A good check point set should
yield a well conditioned solution, and sample the solution well spatially. A
simple way which does not yield such a well conditioned set is to select the
points $\mathbf{y}_{l}^{(c)}\in \mathbb{Y}^{(c)}$ on the boundary of box $%
\Omega _{0}$. This is because the corresponding periodic point $\widetilde{%
\mathbf{y}}_{l}^{(c)}$ will then also be located on the box boundary on the
face opposite to $\mathbf{y}_{l}^{(c)}$. In this case the distances from the
box center to $\mathbf{y}_{l}^{(c)}$ and to $\widetilde{\mathbf{y}}%
_{l}^{(c)} $ will be the same, so both points will be located on a sphere of
radius $r_{l}^{(c)}$. If the basis is based on spherical functions, then
several of these will take the same value for symmetrical points on the
sphere, and the fitting equations may be rank deficient.

To avoid this degeneracy, the set $\mathbb{Y}^{(c)}$ was chosen from points
on the surface of ball $S_{0}$. Three distributions were tried (see Fig. (%
\ref{CheckPoints}): i) random uniformly distributed points, ii) Gaussian
nodes (zeros of the Legendre polynomial along $\theta$, $P_{p}\left( \cos
\theta \right) $, see \cite{Abramowitz:Book1964}) and equispaced with
respect to $\varphi $), and, iii) the almost uniform distribution of points
over the sphere obtained by solving the so-called Thomson problem of the
equilibrium position of mutually repelling electrons constrained to be on
the surface of the sphere \cite{Thomson:PhilMag1904}, see \cite
{Fliege:IMAJNA1999}. Methods ii) and iii) show better results than the
random distribution, as shown in Section 3. Further, we fond that using the
Thomson points for the interpolation in Eq. \ref{5} provides good accuracy.

\subsection{Periodization algorithm}

The algorithm has two parts. In a first preliminary set-up step, denoted
``set'', the check points are determined, and the matrix decompositions
necessary to compute the least squares fit with the matrix $\mathbf{A}$ in
Eq. (\ref{10}) are precomputed. In simulations where the domain $\Omega _{0}$
is fixed and the particles move, as in molecular dynamics, this matrix does
not change, and the cost of the ``set'' step is amortized over the entire
simulation. The second part of the algorithm, denoted ``get,'' computes
the right hand side and solution of the fitting equations via an inexpensive
step such as backsubstitution. The accuracy depends on the choice of basis
functions, and the parameters $P$,$L$, and $R_{b}$. In the next section we
provide both a theoretical and an empirical study for the case of the
Coulombic kernel (Green's function of Laplace's equation).

\subsubsection{Algorithm ``set''}

\begin{enumerate}
\item  Set the circumsphere radius $R_{0}=\frac{1}{2}\sqrt{%
d_{1}^{2}+d_{2}^{2}+d_{3}^{2}}$. Based on the required accuracy determine $%
R_{b}$, $P$, and $L\geqslant P$.

\item  Generate $L$ check points distributed over the surface of the ball $%
S_{0}$, $\mathbf{y}_{l}^{(c)}\in \partial S_{0}$, $l=1,...,L.$ Denote this
point set as $Y_{c1}$.

\item  For each point $\mathbf{y}_{l}^{(c)}=%
\left(y_{l1}^{(c)},y_{l2}^{(c)},y_{l3}^{(c)}\right) ,$ $l=1,...,L,$ find a
point $\widetilde{\mathbf{y}}_{l}^{(c)}\in S_{0}$, such that two Cartesian
coordinates of $\widetilde{\mathbf{y}}_{l}^{(c)}$ are the same as those of
the respective coordinates of $\mathbf{y}_{l}^{(c)}$, while the other
coordinate, $\widetilde{y}_{lk}^{(c)}$ is shifted by $d_{k}$ with respect to 
$y_{lk}^{(c)}$. Denote $Y_{c2}$ $=\left\{ \widetilde{\mathbf{y}}%
_{l}^{(c)}\right\} .$

\item  Form the $L\times P$ fitting matrix $\mathbf{A=}\left\{
A_{lt}\right\} $, $A_{lt}=R_{t}\left(\mathbf{y}_{l}^{(c)}\right)
-R_{t}\left( \widetilde{\mathbf{y}}_{l}^{(c)}\right) $, $l=1,...,L,$ $%
t=1,...,P,$ where $R_{t}\left( \mathbf{y}\right) $ are the basis functions
at the checkpoints.

\item  Compute matrix decomposition of $\mathbf{A}$ necessary to solve the
least squares problem.

\item  (optional) Precompute other parameters which do not depend on the
source distribution. If the summation needs certain auxiliary computations
to ensure convergence, do those steps. For the Coulomb kernels this may
involve computation of the integrals of the basis functions over the box $%
\Omega _{0}$.
\end{enumerate}

\subsubsection{Algorithm ``get''}

\begin{enumerate}
\item  Periodically extend the source box $\Omega _{0}$ to cover the ball $%
S_{b}$ of radius $R_{b}$. The newly generated sources and charges will have
coordinates $X_{\mathbf{p}}=\left\{\mathbf{x}_{i}+\mathbf{p}\right\} $ for
the values of the periodization vector $\mathbf{p}_{1},...,\mathbf{p}_{b}$
from set $\mathbb{P}$ (see Eq. (\ref{2})) and charges $Q_{\mathbf{p}%
}=\left\{ q_{i}\right\} .$ Denote the set of all sources as $X_{b}=X_{0}\cup
X_{\mathbf{p}_{1}}\cup ...\cup X_{\mathbf{p}_{b}}$. These constitute $\Omega
_{b}$.

\item  Find the set of $N_{b}$ sources $X_{near}$ by removing sources from
the set $X_{b}$ that are outside the sphere $S_b$, and satisfy $\left| 
\mathbf{x}_{j}\right| >R_{b}$ $\left( X_{near}=\left\{ \mathbf{x\in }%
X_{b}:\left| \mathbf{x}\right| \leqslant R_{b}\right\} \right) .$

\item  Using the FSA compute $\phi _{near}$ for a given set of evaluation
points $Y_{0}=\left\{ \mathbf{y}_{j}\right\} ,j=1,...,M$ residing in $\Omega
_{0}$ and belonging to sets $Y_{c1}$ and $Y_{c2}$, i.e. for points from the
set $Y=\left\{ \mathbf{y\in }Y_{0}\cup Y_{c1}\cup Y_{c2}\right\} .$ If
gradient computations are needed, compute $\nabla \phi _{near}$ at $\mathbf{y%
}_{j}\in Y_{0}.$

\item  Form the right hand side of the periodization equation $\mathbf{f}$ $%
=\left\{ f_{l}\right\} ,$ $f_{l}=\phi _{near}\left( \widetilde{\mathbf{y}}%
_{l}^{(c)}\right) -\phi _{near}\left( \mathbf{y}_{l}^{(c)}\right) $, $%
l=1,...,L,$ organized in a column vector.

\item  Using the matrix decompositions in the ``set'' step solve the fitting
equations for the $P$ expansion coefficients $\mathbf{C=}\left(
C_{1},...,C_{P}\right) ^{T}.$ This step can formally be written as $\mathbf{C%
}=$ $\mathbf{A}^{\dagger}\mathbf{f.}$

\item  (optional) Compute the constant shift or other modification of the
far field potential, if needed.

\item  Evaluate $\phi _{far}\left( \mathbf{y}_{j}\right) $, and if gradient
computation is needed, $\nabla\phi _{far}\left( \mathbf{y}_{j}\right) $, at $%
\mathbf{y}_{j}\in Y_{0}$.

\item  Get the periodized solution of the problem, $\phi \left(\mathbf{y}%
_{j}\right)=\phi_{near}\left(\mathbf{y}_{j}\right)+\phi_{far}\left(\mathbf{y}%
_{j}\right),$ $\mathbf{y}_{j}\in Y_{0},$ and if gradient computation is
needed, $\nabla \phi \left( \mathbf{y}_{j}\right)$ $=\nabla \phi_{near}\left(%
\mathbf{y}_{j}\right)+$ $\nabla \phi_{far}\left( \mathbf{y}_{j}\right) ,$ $%
\mathbf{y}_{j}\in Y_{0}$.
\end{enumerate}

\begin{remark}
In molecular dynamics and other N-body simulations the source and evaluation
points are the same, so $Y_{0}=X_{0}.$
\end{remark}

\begin{remark}
In Step 5 for the Laplacian kernel and spherical basis functions $\mathbf{C=}
\left( C_{2},...,C_{P}\right) ^{T}$. In this case optional Step 6 provides $%
C_{1}$ (see Eqs (\ref{pfx1})-(\ref{pfx4})). Otherwise set $C_{1}=0$.
\end{remark}

\begin{remark}
Step 2 reduces $\Omega_b$ to the ball $S_b$, which is not necessary, but is
efficient.
\end{remark}

\subsubsection{ Complexity}

Evaluation of the kernel at a single point requires $O(1)$ operations. So, $%
O(P)$ operations are needed to evaluate all basis functions. The complexity, 
$O(P)$, is also achieved when basis functions at a point can be computed
recursively (as for the spherical basis functions). With $L=O(P)$ we can
estimate the complexity of the ``set'' part of the algorithm as 
\begin{equation}
C^{(set)}=O(1)+O\left( P\right) +O(P)+O\left( P\right) +O\left( P^{3}\right)
+\left[ O(P^{5/2})\right] =O\left( P^{3}\right) ,  \label{C1}
\end{equation}
where we assumed that computation of the matrix decomposition (e.g., via QR) 
$O\left( P^{3}\right) $ operations, as $P\sim L$. The term in the square
brackets is the cost of the optional step for the Laplace kernel and
spherical basis functions, using the method presented in Appendix B.

For the ``get'' part of the algorithm, assuming that $N_{b}=O(N)$ and $P\ll N
$ we have 
\begin{equation}
C^{(get)}=O(N)+O(N)+C^{(FSA)}+O\left( P\right) +O(P^{2})+\left[ O\left(
N\right) \right] +O\left( PN\right) +O(N)=O\left( PN\right) +C^{(FSA)},
\label{C2}
\end{equation}
where $C^{(FSA)}$ is the cost of the finite summation algorithm and the cost
of the optional step of the algorithm is put in the square brackets (see
Appendix C for this step for the Laplacian kernel). This cost for the brute
force summation is $O(N^{2})$, while if the FMM is used as the FSA it can be
estimated as follows.

In the FMM, generation of the data structure for $M\sim N$ points is $%
O(Nl_{\max })$ which for deep trees, $l_{\max }=O(\log N)$, results in
formal $O(N\log N)$ complexity. However, in practice the depth of the trees
in three dimensions is relatively small (e.g. $l_{\max }<10$) for sizes $%
N<10^{7}$ and even at larger $l_{\max }$ this cost is much smaller than the
cost of the run part of the FMM. Moreover, the translation time in the FMM
usually dominates over the time of generation and evaluation of expansions
of complexity $O(P_{FMM}N)$, where $P_{FMM}$ is the size of the expansions
in the FMM. For optimal $l_{\max }$ the FMM using $O\left( P_{FMM}^{2\alpha
}\right) $ methods for translations in three dimensions scale as $%
O(P_{FMM}^{\alpha }N)$ or $O(P_{FMM}^{\alpha }N)$, where $\frac{1}{2}%
\leqslant \alpha \leqslant \frac{3}{4}$ for well studied kernels such as
those for the Laplace and Helmholtz equations (in the latter case additional 
$\log ^{\beta }N$, $\beta >0$ factors appear in the algorithm complexity ,
which we drop in the present estimate). Optimized kernel independent FMM has
complexity $O(P_{FMM}N)$. This can be summarized as 
\begin{equation}
C^{(get)}=O\left( P_{FMM}^{\alpha }N\right) +O\left( PN\right) ,\quad \frac{1%
}{2}\leqslant \alpha \leqslant 1.  \label{C3}
\end{equation}

Note that for the Laplacian kernel in three dimensions truncation numbers $%
p=P^{1/2}$ and $p_{FMM}=P_{FMM}^{1/2}$ should increase as $O(\log N)$ for a
fixed absolute $L_{\infty }$-norm error (see error bounds below), while they
are constant for the relative $L_{2}$-norm errors.

\section{Laplacian kernel}

A fundamental feature of the Laplace equation that any constant is a
solution. Thus, one of the basis functions is a constant (say, $R_{1}\left( 
\mathbf{y}\right) \equiv 1$). Moreover, any constant satisfies periodic
boundary conditions, and so this part of the solution cannot be determined,
as $R_{1}\left( \mathbf{y}\right) $ belongs to the null-space of the
Laplacian operator. System (\ref{8}) also shows that $A_{l1}=R_{1}\left( 
\mathbf{y}_{l}^{(c)}\right) -R_{1}\left( \widetilde{\mathbf{y}}%
_{l}^{(c)}\right) =0$ for any $l$, so for any $L\geqslant P$ the rank of
matrix $\mathbf{A}$ cannot exceed $L-1$, in which case the coefficient $%
C_{1} $ can be arbitrary. To remove this rank deficiency of $\mathbf{A}$ we can simply remove the constant basis function from consideration, formulate the problem as the problem of determination of coefficients $C_{t}$ for $t=2,...,L$ and then add an arbitrary constant $C_{1}$ to the solution. Indeed, in many cases the value of the potential is not important, as only differences and gradients determine physical quantities such as the electric field, velocities, forces, etc. For comparison with the FFT-based Poisson equation solutions however, it is desirable to obtain $C_{1}$, in which case an additional condition, zero period average, needs to be imposed. These will be discussed in a separate subsection. Here we just mention that many other equations have the same problem (e.g. the biharmonic equation) and the
null-space of some equations, such as the Helmholtz equation, may have larger dimension, and an analysis similar to the one for the Laplace
equation presented here would be needed.

\subsection{Spherical basis functions}
While the basis functions can be simply selected according to Eq. (\ref{5}), we instead consider the closely related polynomial basis, in which case we can establish error bounds. In spherical coordinates ($r,\theta ,\varphi $) related to the Cartesian coordinates via 
\begin{equation}
x=r\sin \theta \cos \varphi ,\quad x=r\sin \theta \sin \varphi ,\quad
z=r\cos \theta ,  \label{0}
\end{equation}
the local and multipole solutions of the Laplace equation in 3D can be
represented as 
\begin{equation}
R_{n}^{m}\left(\mathbf{r}\right)=\alpha_{n}^{m}r^{n}Y_{n}^{m}(\theta,\varphi),\quad S_{n}^{m}\left( \mathbf{r}\right)
=\beta_{n}^{m}r^{-n-1}Y_{n}^{m}(\theta ,\varphi ),\quad n=0,1,\ldots ,\quad
m=-n,...,n.  \label{1.1}
\end{equation}
Here $R_{n}^{m}\left( \mathbf{r}\right) $ are the regular (local) spherical
basis functions and $S_{n}^{m}\left( \mathbf{r}\right) $ the singular (or
multipole) spherical basis functions; $\alpha _{n}^{m}$ and $\beta_{n}^{m}$
are normalization constants which can be selected by convenience, and $Y_{n}^{m}(\theta ,\varphi )$ are the orthonormal spherical harmonics: 
\begin{align}
Y_{n}^{m}\left( \theta ,\varphi \right) & =N_{n}^{m}P_{n}^{\left| m\right|
}(\mu )e^{im\varphi },\quad \mu =\cos \theta ,\quad  \label{1.2} \\
N_{n}^{m}& =(-1)^{m}\sqrt{\frac{2n+1}{4\pi }\frac{(n-\left| m\right| )!}{%
(n+\left| m\right| )!}},\quad n=0,1,2,...,\quad m=-n,...,n,  \notag
\end{align}
where $P_{n}^{\left| m\right| }\left( \mu \right) $ are the associated
Legendre functions \cite{Abramowitz:Book1964}. We will use the definition of the associated Legendre function $P_{n}^{m}\left( \mu \right) $ that is consistent with the value on the cut $(-1,1)$ of the hypergeometric function $P_{n}^{m}\left( z\right) $ (see Abramowitz \& Stegun, \cite{Abramowitz:Book1964}). These functions can be obtained from the Legendre polynomials $P_{n}\left( \mu \right) $ via the Rodrigues' formula 
\begin{equation}
P_{n}^{m}\left( \mu \right) =\left( -1\right) ^{m}\left( 1-\mu ^{2}\right)
^{m/2}\frac{d^{m}}{d\mu ^{m}}P_{n}\left( \mu \right) ,\quad P_{n}(\mu) =\frac{1}{2^{n}n!}\frac{d^{n}}{d\mu ^{n}}\left( \mu ^{2}-1\right)
^{n}.  \label{1.3}
\end{equation}
Straightforward computation of these basis functions involves several
relatively costly operations with special functions and use of spherical
coordinates. Further, as defined above, these functions are complex, which is an unnecessary expense for real valued computations. In \cite
{Gumerov:JCP2008} real basis functions were defined as 
\begin{equation}
\widetilde{R}_{n}^{m}=\left\{ 
\begin{array}{c}
\func{Re}\left\{ R_{n}^{m}\right\} ,\quad m\geqslant 0 \\ 
\func{Im}\left\{ R_{n}^{m}\right\} ,\quad m<0
\end{array}
\right. ,\quad \widetilde{S}_{n}^{m}=\left\{ 
\begin{array}{c}
\func{Re}\left\{ S_{n}^{m}\right\} ,\quad m\geqslant 0 \\ 
\func{Im}\left\{ S_{n}^{m}\right\} ,\quad m<0
\end{array}
\right. ,  \label{1.4}
\end{equation}
with $R_{n}^{m}$ and $S_{n}^{m}$ defined via Eq. (\ref{1.1}) are 
\begin{align}
\alpha _{n}^{m}& =\left( -1\right) ^{n}\sqrt{\frac{4\pi }{\left( 2n+1\right)
\left( n-m\right) !(n+m)!}},\quad \beta _{n}^{m}=\sqrt{\frac{4\pi \left(
n-m\right) !(n+m)!}{2n+1}},  \label{1.4.1} \\
n& =0,1,....,\quad m=-n,...,n.  \notag
\end{align}
Only local basis functions are needed here, and can be computed via an
efficient recursive process, without spherical coordinates, 
\begin{gather}
\widetilde{R}_{0}^{0}=1,\quad \widetilde{R}_{1}^{1}=-\frac{1}{2}x,\quad 
\widetilde{R}_{1}^{-1}=\frac{1}{2}y,  \label{1.6} \\
\widetilde{R}_{\left| m\right| }^{\left| m\right| }=-\frac{\left( x%
\widetilde{R}_{\left| m\right| -1}^{\left| m\right| -1}+y\widetilde{R}%
_{\left| m\right| -1}^{-\left| m\right| +1}\right) }{2\left| m\right| }%
,\quad \widetilde{R}_{\left| m\right| }^{-\left| m\right| }=\frac{\left( y%
\widetilde{R}_{\left| m\right| -1}^{\left| m\right| -1}-x\widetilde{R}%
_{\left| m\right| -1}^{-\left| m\right| +1}\right) }{2\left| m\right| }%
,\quad \left| m\right| =2,3,....  \notag \\
\widetilde{R}_{\left| m\right| +1}^{m}=-z\widetilde{R}_{\left| m\right|
}^{m},\quad m=0,\pm 1,....,  \notag \\
\widetilde{R}_{n}^{m}=-\frac{(2n-1)z\widetilde{R}_{n-1}^{m}+r^{2}\widetilde{R%
}_{n-2}^{m}}{\left( n-\left| m\right| \right) (n+\left| m\right| )},\quad
n=\left| m\right| +2,....,\quad m=-n,...,n.  \notag
\end{gather}

While this basis is good for the FMM, the matrix $\mathbf{A}$ for fitting $\phi_{far}$ was found to be poorly conditioned, because the functions decay strongly. We fix this problem using the following renormalized basis 
\begin{equation}
\widehat{R}_{n}^{m}=\sqrt{\left( n-m\right) !(n+m)!}\widetilde{R}%
_{n}^{m},\quad n=0,1,....,\quad m=-n,...,n.  \label{1.7}
\end{equation}
This basis can be obtained in the same manner as $\left\{\widetilde{R}%
_{n}^{m}\right\}$ was from the complex basis $R_{n}^{m}$, where $\alpha
_{n}^{m}=$ $\left( -1\right) ^{n}\sqrt{4\pi /\left( 2n+1\right) }$. Complex basis functions $R_{n}^{m}$ with a similar normalization $\alpha _{n}^{m}$ (without the factor $\left( -1\right) ^{n}$) were used in \cite{Epton:SIAMJSC1995}. Note further that the $p$-truncated expansion of a harmonic function $\phi_{far} $ over the basis (\ref{1.7}) can be written as 
\begin{eqnarray}
\phi_{far}\left(\mathbf{y}\right)&=&\sum_{n=0}^{p-1}\sum_{m=-n}^{n} \widehat{C}_{n}^{m}\widehat{R}_{n}^{m} \left( \mathbf{y}\right)=%
\sum_{t=1}^{P}C_{t}R_{t}\left( \mathbf{y}\right),\quad C_{t}=\widehat{C}%
_{n}^{m},\;\; R_{t}\left( \mathbf{y}\right)=\widehat{R}_{n}^{m}\left( 
\mathbf{y}\right), \;\; P=p^{2},  \notag \\
t &=&(n+1)^{2}-(n-m),\quad n=0,1,....,p-1,\quad m=-n,...,n.  \label{1.8}
\end{eqnarray}
The latter form of the sum, where stacking of coefficients is used, is
consistent with (\ref{6}). The gradient of the potential is needed to
compute the force. This can be computed as 
\begin{equation}
\nabla \phi _{far}\left( \mathbf{y}\right) =\sum_{n=1}^{p-1}\sum_{m=-n}^{n}\widehat{\mathbf{E}}_{n}^{m}\widehat{R}_{n}^{m}\left(\mathbf{y}\right)
=\sum_{t=2}^{P}\mathbf{E}_{t}R_{t}\left( \mathbf{y}\right) ,  \label{1.9}
\end{equation}
where $\widehat{\mathbf{E}}_{n}^{m}$ are vectors in $\mathbb{R}^{3}$. In \cite{Gumerov:JCP2013} one can find relations between the coefficients of the potential and the gradient.

\subsection{Error bounds}
For the Laplacian kernel $K$, Eq. (\ref{2}), $p$-truncated expansions (\ref{4}) over the basis $\left\{ R_{n}^{m}\right\} $, Eq. (\ref{1.1}), have a
well-known error bound 
\begin{equation}
\left| \epsilon _{j}^{(P)}\right| <\frac{1}{\left| \mathbf{x}_{j}\right|
-R_{0}}\left( \frac{R_{0}}{\left| \mathbf{x}_{j}\right| }\right) ^{p},\quad
p=P^{1/2}.  \label{errb1}
\end{equation}
The expansion error (\ref{6}) due to all sources located in $\mathbb{R}^{3}\backslash \Omega _{b}$ then can be bounded as 
\begin{eqnarray}
\left| \epsilon ^{(P)}\right| &<&\frac{\max \left| q_{i}\right| }{R_{b}-R_{0}%
}\sum_{\mathbf{x}_{j}\notin \Omega _{b}}\left( \frac{R_{0}}{\left| \mathbf{x}%
_{j}\right| }\right) ^{p}\leqslant \frac{\max \left| q_{i}\right| }{%
R_{b}-R_{0}}\sum_{\mathbf{x}_{j}\in \mathbb{R}^{3}/\Omega _{b}}\left( \frac{%
R_{0}}{\left| \mathbf{x}_{j}\right| }\right) ^{p}  \label{errb2} \\
&\leqslant&\frac{\max \left| q_{i}\right| }{R_{b}-R_{0}}\int_{\mathbb{R}%
^{3}/\Omega _{b}}n\left( \mathbf{x}\right) \left( \frac{R_{0}}{r}\right)
^{p}dV,\quad  \notag \\
n\left( \mathbf{x}\right) &=&\sum_{\mathbf{x}_{j}\in \mathbb{R}^{3}/\Omega
_{b}}\delta \left( \mathbf{x}-\mathbf{x}_{j}\right) ,\quad r=\left| \mathbf{x%
}\right| .  \notag
\end{eqnarray}
Here we introduced the the number density $n \left( \mathbf{x}\right) $, which for integral estimates can be replaced with a constant density $n_{0}=N/V_{0}$, where $V_{0}$ is the volume of box $\Omega _{0}$, $V_{0}=d_{1}d_{2}d_{3}$. In this case the integral can be evaluated as 
\begin{equation}
\int_{\mathbb{R}^{3}/\Omega _{b}}n\left( \mathbf{x}\right) \left( \frac{R_{0}}{r}\right) ^{p}dV\sim 4\pi n_{0}\int_{R_{b}}^{\infty }\left( \frac{R_{0}}{r}\right) ^{p}r^{2}dr=\frac{4\pi n_{0}R_{b}^{3}}{p-3}\lambda ^{-p},\quad
\lambda =\frac{R_{b}}{R_{0}}.  \label{errb3}
\end{equation}
Hence, we have an approximate error bound 
\begin{equation}
\left| \epsilon ^{(P)}\right| \lesssim \frac{4\pi n_{0}R_{b}^{3}}{R_{b}-R_{0}
}\frac{\max \left| q_{i}\right| }{p-3}\lambda ^{-p}.  \label{errb4}
\end{equation}
For a cubic domain we have $d_{1}=d_{2}=d_{3}=d,$ $R_{0}=\frac{1}{2}d\sqrt{3}%
,$ $R_{b}=\lambda R_{0}$, and we get 
\begin{equation}
\left| \epsilon ^{(P)}\right| \lesssim \frac{3\pi N\max \left| q_{i}\right| 
}{d}\frac{1}{\left( \lambda -1\right) \left( p-3\right) \lambda ^{p-3}}.
\label{errb4.1}
\end{equation}
The actual error achieved in practice is expected to be much smaller than this estimate since it neglects cancellation effects due to the total charge neutrality. Also in the above equation one can set $d=1$ to obtain a non-dimensional measure of the absolute error (since the Laplace equation is scale-independent).

\subsection{Optimization when using the FMM}
There are three free parameters, $p$, $p_{FMM}$, and $\lambda $, which can be selected to optimize algorithm performance. Assuming that the number of charges, their intensities and distribution as well as the domain $\Omega_{0}$ are fixed, and computations performed with some prescribed tolerance, $\epsilon $, the optimization problem can be formulated as 
\begin{equation}
\epsilon _{P}\left( p,\lambda \right) =\epsilon ,\quad \epsilon _{FMM}\left(p_{FMM},\lambda \right) =\epsilon ,\quad C^{(get)}\left( p,p_{FMM},\lambda\right) \rightarrow \min ,  \label{o1}
\end{equation}
where $\epsilon _{P}$ and $\epsilon _{FMM}$ are the error bounds for the periodization and the FMM respectively, while $C^{(get)}$ is the cost of the ``get'' step. In practice, these performance dependences should be determined experimentally, using the qualitative theoretical estimates provided below for guidance. Our tests show that the cost of the FMM is the major contributor to the overall cost of the algorithm.

We set the parameter $p_{FMM}$ by the prescribed accuracy $\epsilon $, and approximate $\epsilon _{P}\left( p,\lambda \right) $ as 
\begin{equation}
\epsilon _{P}\left( p,\lambda \right) =B_{P}\lambda ^{-p},\quad p\left(
\lambda \right) =\frac{\ln \left( B_{P}/\epsilon \right) }{\ln \lambda },
\label{o2}
\end{equation}
where $B_{P}$ is some constant. The number of evaluation points is $%
M_{b}=N+2L$, while the number of sources to be summed is $%
N_{b}=A_{gf}\lambda ^{3}N$, where $A_{gf}$ is a geometric factor,
\begin{equation}
A_{gf}=\frac{\pi }{6}\frac{\left( d_{1}^{2}+d_{2}^{2}+d_{3}^{2}\right) ^{3/2}%
}{d_{1}d_{2}d_{3}}.  \label{o2.1}
\end{equation}
For a perfectly optimized FMM the complexity estimate is provided in
Appendix A, Eq. (\ref{F4}), where the ratio of the domains occupied by the sources and evaluation points is $V_{s}/V_{r}=B_{gf}\lambda ^{3}$, where $1\leqslant $ $B_{gf}\leqslant A_{gf}$ and $B_{gf}$ tends to 1 and $A_{gf}$ at relatively low and high levels of subdivision, $l_{\max}$,
respectively. This is due to the fact that the check point set is
distributed over the surface of a sphere, and they should provide a negligible contribution to the overall complexity at large $l_{\max }$. Assuming $L\sim 2p^{2}$ the cost can be estimated as 
\begin{eqnarray}
C^{(FMM)} &=&C_{td}\sqrt{\frac{A_{gf}}{B_{gf}}MN}+C_{gen}A_{gf}N\lambda
^{3}+C_{ev}M,  \label{o3} \\
M &=&N\left( 1+\frac{4p^{2}\left( \lambda \right) }{N}\right) .  \notag
\end{eqnarray}
It is seen that the complexity is a sum of the decreasing and increasing
functions, a minimum at some $\lambda =\lambda _{opt}$ \ is expected. It is not difficult to find it for the case when $\lambda $ is close to $1$ and $4p^{2}\ll N.$ Indeed, introducing $\xi =\lambda -1\ll 1$ we have from Eq. (\ref{o2}) $p\sim \xi ^{-1}\ln \left( B_{P}/\epsilon \right) $ and expanding Eq. (\ref{o3}) at small $\xi $ and $p^{2}/N$ we obtain 
\begin{equation}
C^{(FMM)}=N\left[ C_{td}\left( \frac{A_{gf}}{B_{gf}}\right)
^{1/2}+C_{gen}A_{gf}+C_{ev}+\frac{2\ln ^{2}\left( B_{P}/\epsilon \right) }{N\xi ^{2}}\left( C_{td}\left( \frac{A_{gf}}{B_{gf}}\right)
^{1/2}+2C_{ev}\right) +3C_{gen}A_{gf}\xi \right] .  \label{o4}
\end{equation}
This function has a minimum at 
\begin{equation}
\xi _{opt}=\left[ \frac{4\ln ^{2}\left( B_{P}/\epsilon \right) }{%
3C_{gen}A_{gf}N}\left( C_{td}\left( \frac{A_{gf}}{B_{gf}}\right)
^{1/2}+2C_{ev}\right) \right] ^{1/3}.  \label{o5}
\end{equation}
which shows that at large $N$ optimal $\lambda $ should be shifted towards the limit $\lambda =1.$ Also this shows that at large $N$ we have $p^{2}/N\sim N^{-1/3}\rightarrow 0$, which justifies the above asymptotic solution, and shows that the part of the overall cost depending on $\lambda $ in Eq. (\ref{o4}) tends to zero. 

\subsection{Constant shift in potential}
Several methods can be proposed to determine coefficient $C_{1}$ if it is
needed. In particular because we choose to compare our results with the
Ewald summation method, we need it. Of course, the simplest case is that
when the potential value, $\phi _{0}$, is prescribed or known at some point $%
\mathbf{y}_{0}$, in which case 
\begin{equation}
C_{1}=\phi _{0}-\phi _{near}\left( \mathbf{y}_{0}\right)
-\sum_{t=2}^{P}C_{t}R_{t}\left( \mathbf{y}_{0}\right) .  \label{pfx1}
\end{equation}

Note then that the Fourier based methods for periodization of Green's
function produce solution with some mean of the potential $\phi _{mean}$
(since the zero mode of the Fourier transform is zeroed). This particular
solution corresponds to 
\begin{equation}
\left\langle \phi \right\rangle _{\Omega _{0}}=\frac{1}{V_{0}}\int_{\Omega
_{0}}\phi \left( \mathbf{y}\right) dV\left( \mathbf{y}\right) =\phi _{mean}.
\label{pfx2}
\end{equation}
In this case for consistency we should set 
\begin{equation}
C_{1}=\phi _{mean}-\frac{1}{V_{0}}\int_{\Omega _{0}}\phi _{near}\left( 
\mathbf{y}\right) dV\left( \mathbf{y}\right)
-\sum_{t=2}^{P}C_{t}R_{t}^{(0)},\quad R_{t}^{(0)}=\frac{1}{V_{0}}%
\int_{\Omega _{0}}R_{t}\left( \mathbf{y}\right) dV\left( \mathbf{y}\right) .
\label{pfx3}
\end{equation}
As shown in Appendix D, the Ewald summation produces $\phi _{mean}=0$, and
we do the same for our method. Integrals $R_{t}^{(0)}$ can be computed
relatively easy, since $R_{t}\left( \mathbf{y}\right) $ are polynomials in
the Cartesian coordinates of $\mathbf{y}$ of degree which does not exceed $p-1$, for which case exact quadratures exist. In fact, for a given box size ratio (e.g. for cube) these can be precomputed, scaled and used
independently of particular source distribution (see Appendix B). The first
integral can be represented as a sum 
\begin{equation}
\frac{1}{V_{0}}\int_{\Omega _{0}}\phi _{near}\left( \mathbf{y}\right)
dV\left( \mathbf{y}\right) =\sum_{\mathbf{x}_{j}\in \Omega _{b}}q_{j}\Phi
_{0}\left( \mathbf{x}_{j}\right) ,\quad \Phi _{0}\left( \mathbf{x}\right) =%
\frac{1}{V_{0}}\int_{\Omega _{0}}K\left( \mathbf{y}-\mathbf{x}\right)
dV\left( \mathbf{y}\right) .  \label{pfx4}
\end{equation}
In Appendix C we provide analytical expressions for functions $\Phi
_{0}\left( \mathbf{x}\right) $. Despite their unwieldiness, their
computation for a given $\mathbf{x}$ is $O(1)$. Overall it is a $O\left(
N\right) $ procedure to compute the sum and constant $C_{1}$, which is
consistent with the overall complexity of the method. There also exist
symmetries for periodic location of sources, which can be used to accelerate
these computations, if this becomes an issue. Note also that results of
Appendix C can be applied for computation of integrals representing the far
field (\ref{pfx3}) in the case when the RBF, Eq. (\ref{5}), is used.

\section{Numerical tests}

To check the accuracy and performance of the method we conducted several
numerical tests. There are very few known analytical solutions, so for
comparison we also implemented and tested a simple version of the Ewald
summation method as an alternative method (see Appendix D).

\subsection{Small size tests}

As validation, we performed tests with different number of sources in the
box. First, we conducted a small size test, with a cubic domain $\Omega _{0}$
and eight sources of charges $q_{i}=\pm 1$ located at the vertices, so that
neighboring sources have opposite charges and the infinite domain forms a
regular equispaced grid. Physically this corresponds to crystal structures,
such as formed by molecules NaCl$.$ As the reference for accuracy tests we
computed the Madelung constant for this crystal \cite{Madelung:PhysZs1918}, 
\begin{eqnarray}
Ma_{\text{Na}} &=&-Ma_{\text{Cl}}=\phi \left( \mathbf{y}_{\text{Na}}\right)
=R_{\text{NaCl}}\sum_{\mathbf{p}}\sum_{i=1}^{N}q_{i}K\left( \mathbf{y}_{%
\text{Na}}-\mathbf{x}_{i}+\mathbf{p}\right)  \label{st1} \\
&=&\sum_{\substack{ j,k,l=-\infty ,  \\ j^{2}+k^{2}+l^{2}\neq 0}}^{\infty }%
\frac{\left( -1\right) ^{j+k+l}}{\left( j^{2}+k^{2}+l^{2}\right) ^{2}}%
=-1.74756459463318219...,  \notag
\end{eqnarray}
where $\mathbf{y}_{Na}$ is the location of Na atom and $R_{\text{NaCl}}$ is
the distance between the closest neighbor atoms (in the tests we used $%
d_{1}=d_{2}=d_{3}=1$, in which case $R_{\text{NaCl}}=0.5$). We also computed
this constant using the Ewald summation and compared spatial distributions
of the potential. As the measures of the relative errors we used 
\begin{equation}
\epsilon _{M}=\left| \frac{Ma^{(comp)}}{Ma^{(true)}}-1\right| ,\quad
\epsilon _{2}=\frac{\left\| \phi ^{(Present)}-\phi ^{(Ewald)}\right\| _{2}}{%
\left\| \phi ^{(Ewald)}\right\| _{2}},\quad \left\| \phi \right\| _{2}=\sqrt{%
\frac{1}{M}\sum_{i=1}^{M}\phi ^{2}\left( \mathbf{y}_{i}\right) },
\label{st2}
\end{equation}
where $\mathbf{y}_{i}\in \Omega _{0}$ are the receivers located on the grid
used for the Ewald summation.

For the high accuracy test, we selected a $44\times 44\times 44$ grid, $\xi
=12$, and the sampling neighborhood for each source $N_{r}=20$ for the Ewald
method (see Appendix D). This setting provides $\epsilon _{M}\approx 10^{-14}
$ (i.e. 14 digits of the Madelung constant). High accuracy test for the
present method was performed with $p=35,$ $R_{b}=1.5$ $\left( \lambda
=R_{b}/R_{0}=\sqrt{3}\right) $, which resullts in errors $\epsilon
_{M}\approx 6\cdot 10^{-14}$ and $\epsilon _{2}\approx 9\cdot 10^{-13}$. For
the middle accuracy test we used $24\times 24\times 24$ grid, $\xi =10$, and
the sampling neighborhood for each source $N_{r}=10$, in which case the
Ewald method results in $\epsilon _{Ma}\approx 6\cdot 10^{-9}$. In our
method we used $p=16,$ $R_{b}=1.5$, which produced errors $\epsilon
_{M}\approx 7\cdot 10^{-8}$ and $\epsilon _{2}\approx 10^{-6}$. These tests
show that errors $\epsilon _{M}$ and $\epsilon _{2}$ are related and the
former one approximately one order of magnitude smaller than the latter. So
in the following accuracy tests we measured only $\epsilon _{M}$ for our
method, which is independent of the Ewald summation routine. These
computations were performed for the check points $\mathbf{y}_{l}^{(c)}$
distributed on the Gauss spherical grid. 
\begin{figure}[tbh]
\begin{center}
\includegraphics[trim=0.5in 0in 0in 0.45in, width=5.5in]
{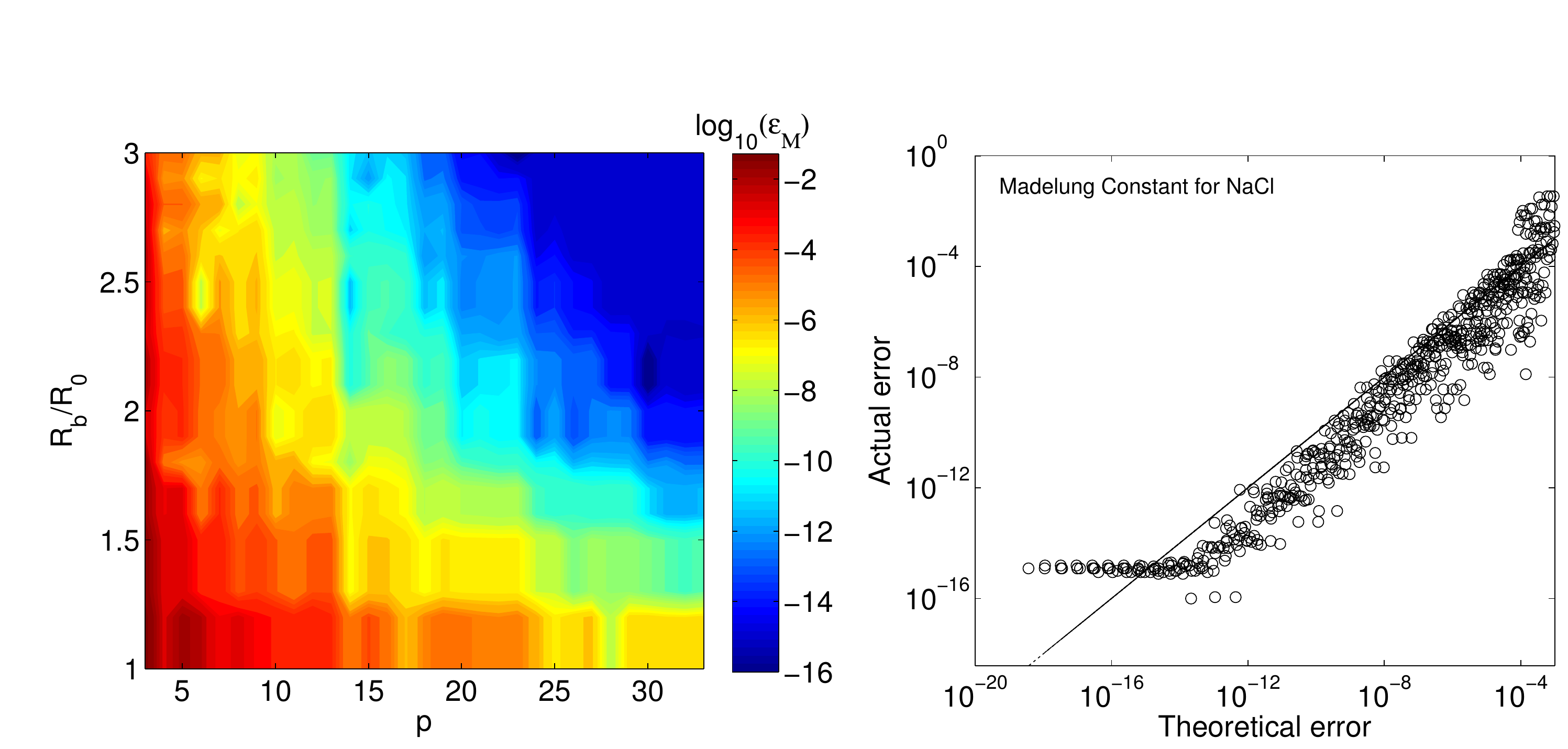}
\end{center}
\caption{The relative error, $\protect\epsilon _{M}$, in computations of the
Madelung constant for NaCl crystal using the present method (colors). The
chart on the graph compares the theoretical computational error, $\protect%
\epsilon _{th}=C_{\protect\epsilon }\left( R_{0}/R_{b}\right) ^{p}$ with the
actual error for all data points used to plot the chart on the left.
Ideally, the graph on the right should be a straight line (shown). Constant $%
C_{\protect\epsilon }$ was set as $\protect\epsilon _{M}/\protect\epsilon
_{th}$ for $p=10$ and $\protect\lambda =R_{b}/R_{0}=1.5$.}
\label{FigErrMadelung}
\end{figure}

Figure \ref{FigErrMadelung} shows the dependence of $\epsilon _{M}$ computed
for 651 values of parameters $\lambda =R_{b}/R_{0}$ and $p$ controlling the
accuracy. The chart on the right shows that the computational errors are
consistent with theoretical error bound $\epsilon _{th}=C_{\epsilon }\left(
R_{0}/R_{b}\right) ^{p}$. For very small values of $\epsilon _{th}$ the
computational errors are affected by the double precision roundoff errors.
This shows that the parameters can be set to achieve the required accuracy. 
{\small 
\begin{table}[htb]
\caption{Error $\protect\epsilon _{M}$ for different check point
distributions at $\protect\lambda =R_{b}/R_{0}=\protect\sqrt{3}.$ }
\label{Table1}
\begin{center}
{\small 
\begin{tabular}{|l|l|l|l|l|l|l|}
\hline
$p$ & Gauss sph. grid & T(256) & T(400) & Rand.$L=2p^{2}$ & Rand.$L=3p^{2}$
& Rand.$L=5p^{2}$ \\ \hline
8 & 4.85(-6) & 3.87(-6) & 3.90(-6) & 3.39(-5) & 2.21(-5) & 1.92-(5) \\ \hline
12 & 1.17(-7) & 4.60(-8) & 6.28(-8) & 1.53(-6) & 1.04(-6) & 5.10(-7) \\ 
\hline
16 & 7.16(-8) & 3.43(-7) & 2.74(-8) & 6.26(-7) & 3.50(-7) & 1.85(-7) \\ 
\hline
\end{tabular}
}
\end{center}
\end{table}
}

Table 1 shows some results of the tests with different distributions of the
check points. Here for the case of random distributions for any set size we
performed 100 runs and the maximum error is reported. It is seen that the
lowest errors were achieved using the Thomson point distributions. The
number of such points should be not less than $p^{2}-1$, which is necessary
(but not sufficient) condition for the use of the present method. When the
number of check points approaches $p^{2}-1$ the accuracy of the method
deteriorates. Conclusion here is that if a database of the Thomson points or
some analogous method of deterministic uniform distribution of the check
points exsist, then that method is recommended. In fact, the Gauss spherical
grid also provides good results (the order of the error is the same). This
grid is easy to generate for any $p$, and that is why this was used in the
tests. The error for random distributions is about one order of magnitude
larger than that for the Gauss spherical grid or for the Thomson points. It
slowly decays with the growing oversampling. Perhaps, there is no reason to
use random sets, which anyway show strong dependence of the error on $p$ and
also can be used if needed.

We also performed the accuracy test for different basis functions (RBF, Eq. (%
\ref{5})), where 256 Thomson sampling sources $\mathbf{x}_{t}^{(s)}$ were
located on ball $S_{b}$, while the check points were the same as for the
last line of Table 1 ($p=16$). In this case we obtained $\epsilon
_{M}\approx 1.32\cdot 10^{-7}$, which is approximately two times larger than
the error when using the spherical basis functions.

\subsection{Large scale tests}

The large scale tests were conducted for systems with $N$ up to $2^{23}\sim 10^{7}$, for which $O(N\log N)$ summation algorithms are needed. The main purpose of these tests was to check the performance and scaling of the present algorithm. The reported wall clock times were measured on an Intel QX6780 (2.8 GHz) 4 core PC with 8 GB RAM and averaged over ten runs of the same case.

We used a well-tested standard version of the FMM for the Laplacian kernel in three dimensions, where all translations are performed with $O\left(p_{FMM}^{3}\right) $ complexity using the rotation-coaxial translation-back rotation (RCR) algorithm. The code implements a standard multipole-to-local translation stencil with the maximum 189 neighborhood (see details in \cite{Gumerov:TR2005}). The code was parallelized for 4 core CPU machine using OpenMP with parallelization efficiency close to 100\%. While faster versions of the FMM were available to the researchers (say utilizing graphics processors, \cite{Gumerov:JCP2008}), the version used for the tests was
selected to provide consistent scaling of different algorithm parts, as the
periodization algorithm was implemented on multicore CPUs. For the tests we treated the FMM as a black box FSA and used it ``as is'' without any modifications.

First we ran accuracy tests, when the charges have random intensity, $\pm 1$, and zero total sum and are located inside a unit cube at regularly spaced
grid points (subgrids of $60\times 60\times 60$ grid). The potential at
charge locations (reference solution) was computed with high accuracy using
the Ewald method (grid $60\times 60\times 60$, $\xi =15,N_{r}=24$). The
present method with $p=40$ and $p_{FMM}=30$ for this case showed error $\epsilon _{2}\approx 1.2\cdot 10^{-10}$. Further performance tests were
conducted with lower tolerance to ensure that the error of the reference
solution does not affect error and optimization studies. In all cases the
Gauss spherical grid ($L=2p^{2}-p$) was used for the check points. 
\begin{figure}[tbh]
\begin{center}
\includegraphics[trim=1.5in 0in 0.5in 0.2in,width=6.0in]
{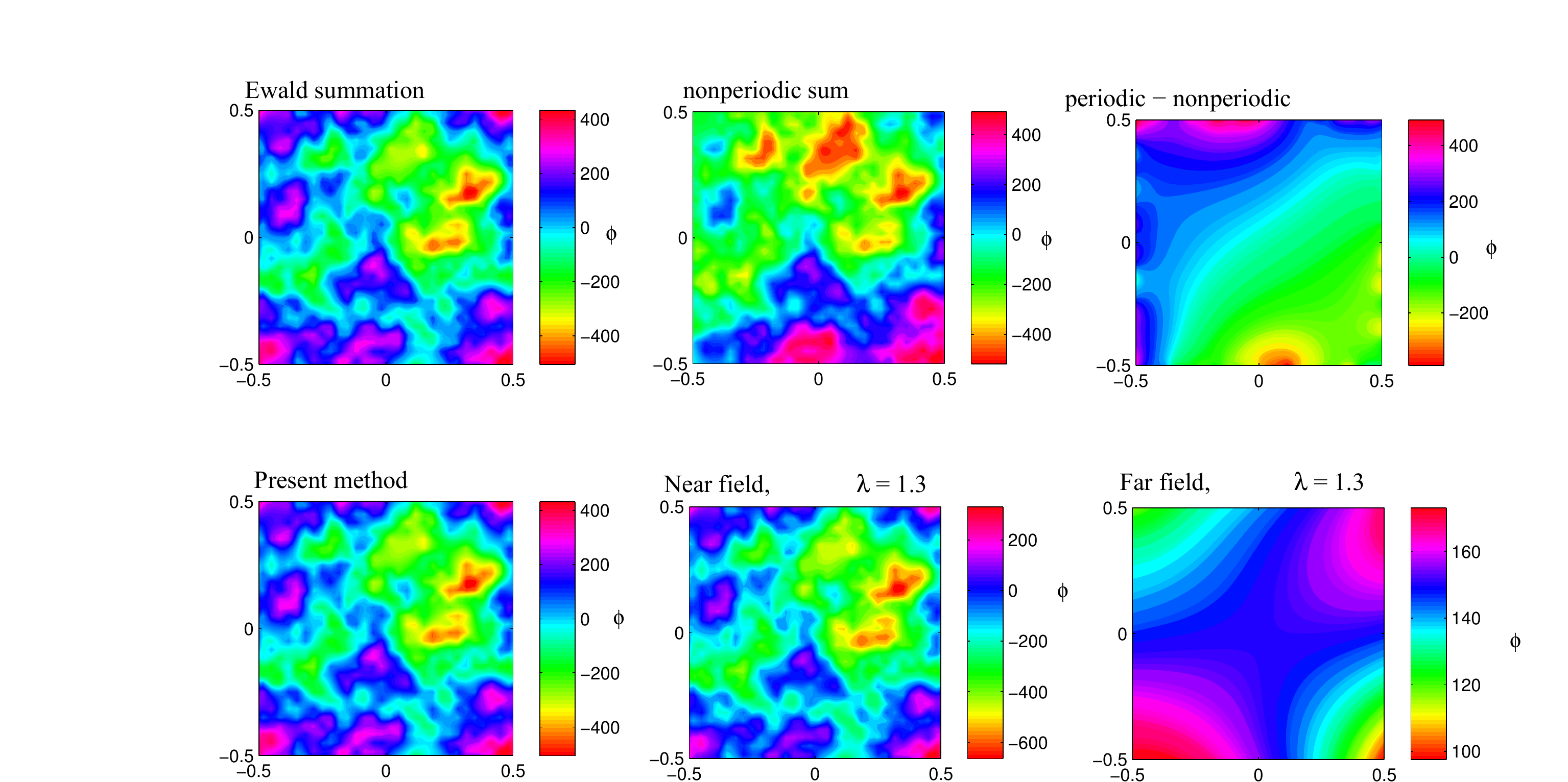}
\end{center}
\caption{ Comparison of periodic solutions at the median plane $z=0$,
obtained by different methods for $30\times 30\times 30$ sources of random
intensity $q_{i}=\pm 1$ in a unit box. The top row shows respectively the
periodic solution obtain by the Ewald method, the non-periodic solution, and
the difference between the former and latter potentials. The bottom row
shows the periodic solution obtained by the method proposed in this paper,
and its near and far field components for $\protect\lambda =1.1$.
Computations performed with $p=80,$ $p_{FMM}=16$.}
\label{FigCompareFields}
\end{figure}

Figure \ref{FigCompareFields} illustrates distribution of potentials
generated by 27,000 charges in a box computed by two different methods
sampled on $60\times 60\times 60$ grid. It is seen that periodic solutions
obtained using the Ewald method and present method are almost the same ($%
\epsilon _{2}<5\cdot 10^{-7}$), while substantially different from the
non-periodic solution (the free field generated by the same sources). This
also shows that accounting for the near field (sources in $\Omega _{b}$)
substantially improves the non-periodic solution, but the far field
component is still of magnitude comparable with the near field. This far
field is smooth and its addition to the near field results in an accurate
periodic solution.

{\small 
\begin{table}[tbh]
\caption{Performance for different $\protect\lambda $ for tolerance $\protect%
\epsilon _{tol,2}=5\cdot 10^{-7}$ at $N=27,000$, $p_{FMM}=16.$ }
\label{Table2}
\begin{center}
{\small 
\begin{tabular}{|l|l|l|l|l|l|l|l|}
\hline
$\lambda $ & $p_{\min }$ & $T^{(get)},$s & $T_{(no\_mean)}^{(get)},$s & $%
T^{(FMM)},$s & $N_{b}$ & $N+2L$ & $T^{(set)},$s \\ \hline
1.1 & 80 & 2.81 & 2.64 & 2.03 & 97,780 & 52,440 & 150 \\ \hline
1.2 & 44 & 1.95 & 1.73 & 1.58 & 126,744 & 34,656 & 5.31 \\ \hline
1.3 & 34 & 2.03 & 1.78 & 1.66 & 161,072 & 31,556 & 1.27 \\ \hline
1.5 & 24 & 2.21 & 1.77 & 1.69 & 247,742 & 29,256 & 0.29 \\ \hline
1.7 & 19 & 2.53 & 1.98 & 1.90 & 360,640 & 28,406 & 0.13 \\ \hline
\end{tabular}
}
\end{center}
\end{table}
} As the theory predicts existence of an optimal value of parameter $\lambda 
$ for a given tolerance we conducted tests to determine this value
experimentally. Tables 2 and 3 display the results of these tests. Here we
computed potential alone (no gradient computations). The difference between
the cases shown in the tables is in the number of charges (27,000 and
216,000, respectively). In these tests $p_{FMM}=16$, which provided the
relative $L_{2}$-norm error of the FMM itself smaller than tolerance $%
\epsilon _{tol,2}$. Since the truncation number $p$ changes discretely,
there is the minimal integer $p=p_{\min }$ at which $\epsilon _{2}<\epsilon
_{2,tol}$, where $\epsilon _{2}$ is defined by Eq. (\ref{st2}). This $p$ is
slightly depends on $N$ and is shown in the tables. The periodization
algorithm was executed with this $p.$ The tables also show the number of
sources, $N_{b}$, and the total number of evaluation points, $N+2L$. These
numbers provide data of the size of the problem solved by the FMM. It is
seen that the FMM execution time is a non-monotonic function of $\lambda $,
as at the increasing $\lambda $ we have increasing $N_{b}=O\left( N\lambda
^{3}\right) $ and decreasing $L\sim 2p_{\min }^{2}$. It is noticeable that
despite of substantial change of $N_{b}$ the FMM time does not change
significantly. The time of the ``set'' part of the algorithm increases
dramatically at large $p$ (as $p^{6}$). However, for a given $p$ and
computational domain the pseudoniverse matrix can be precomputed and stored
independently on the number of charges and their distribution. So this
should not be considered as a limiting factor. These tests bring us to
conclusion that practical optimal $\lambda $ are in the range $\sim 1.2-1.5$%
. Smaller or larger $\lambda $ can be also used based on particular problems
and other issues (e.g. memory complexity).

{\small 
\begin{table}[htb]
\caption{Performance for different $\protect\lambda $ for tolerance $\protect%
\epsilon _{tol,2}=5\cdot 10^{-7}$ at $N=216,000$, $p_{FMM}=16.$ }
\label{Table3}
\begin{center}
{\small 
\begin{tabular}{|l|l|l|l|l|l|l|l|}
\hline
$\lambda $ & $p_{\min }$ & $T^{(get)},$s & $T_{(no\_mean)}^{(get)},$s & $%
T^{(FMM)},$s & $N_{b}$ & $N+2L$ & $T^{(set)},$s \\ \hline
1.1 & 80 & 15.7 & 14.6 & 11.5 & 782,131 & 241,440 & 150 \\ \hline
1.2 & 44 & 12.6 & 11.2 & 10.2 & 1,015,037 & 223,656 & 5.31 \\ \hline
1.3 & 34 & 12.8 & 11.1 & 10.3 & 1,291,247 & 220,556 & 1.27 \\ \hline
1.5 & 25 & 15.0 & 12.3 & 11.7 & 1,983,665 & 218,450 & 0.32 \\ \hline
1.7 & 19 & 18.7 & 14.7 & 14.1 & 2,887,393 & 217,406 & 0.13 \\ \hline
\end{tabular}
}
\end{center}
\end{table}
}

Effect of the basis functions on the algorithm performance is shown in Table
4. As in the small size tests 256 Thomson points were selected as the
centers of the RBFs. Comparison of the performance obtained using the RBFs
vs the spherical basis functions, show that the latter choice is beneficial
in terms of the accuracy, while the speed is approximately the same.
Nonetheless, the errors for both bases are of the same order, while the RBF
implementation is slightly simpler.

{\small 
\begin{table}[tbh]
\caption{Performance for different basis functions at $\protect\lambda =1.5,$
$p=16$, $p_{FMM}=12.$ }
\label{Table4}
\begin{center}
{\small 
\begin{tabular}{|l|l|l|l|l|l|}
\hline
$N$ & Basis & $\epsilon _{2}$ & $T^{(get)},$s & $T_{(no\_mean)}^{(get)},$s & 
$T^{(set)},$s \\ \hline
27,000 & Spherical & 1.3(-5) & 1.67 & 1.30 & 0.08 \\ \hline
& RBF & 2.6(-5) & 1.74 & 1.35 & 0.06 \\ \hline
216,000 & Spherical & 1.6(-5) & 12.3 & 9.57 & 0.08 \\ \hline
& RBF & 2.6(-5) & 12.7 & 9.95 & 0.06 \\ \hline
\end{tabular}
}
\end{center}
\end{table}
}

\begin{figure}[tbh]
\begin{center}
\includegraphics[trim=1.5in 0 0 0, width=6.25in]
{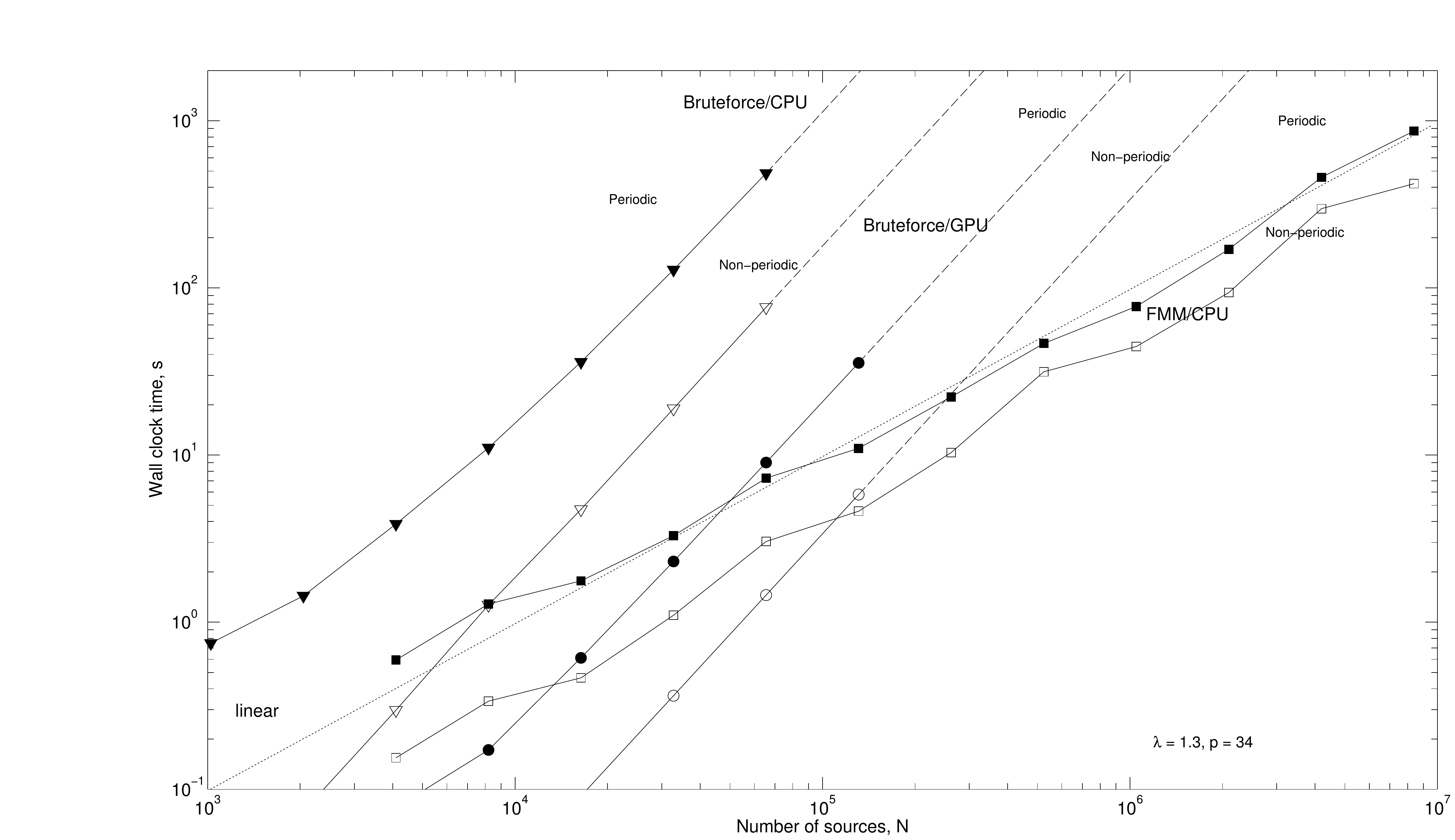}
\end{center}
\caption{Wall clock times of the present algorithm with the FMM as FSA for $%
\protect\lambda =1.3$, $p=34$ (filled squares), and with brute-force 4 core
CPU summation (filled triangles), and Tesla C2050 GPU summation (filled
circles), for a periodic system replicating distribution of $N$ random
sources in a unit cube (potential and gradient computations, Laplacian
kernel in three dimensions). The corresponding empty squares, triangles and
circles, connected by a solid line show performance of the particular FSA
for the non-periodic problem with the same source distribution in the unit
cube and the same truncation number $p_{FMM}=16$. The dashed lines continue
the trendlines to indicate performance, if memory resources had not been
exceeded. The quadratic scaling of the brute-force summation, and the linear
scaling for the FMM are seen. Brute force GPU sums for this case are faster
till about $N\simeq 40,000$.}
\label{Performance}
\end{figure}
Figure \ref{Performance} illustrates scaling of the algorithm with the
problem size. In this test $N$ sources were distributed randomly in the unit
box and both $\phi $ (with zero mean) and $\nabla \phi $ were computed at
the source locations. Here only the time for the ``get'' part is displayed.
For comparison we also plotted the wall clock time for solution of the same
non-periodic problem, where the FMM with the same $p_{FMM\text{ }}$was used.
(The discussion of GPU times is below). While for the non-periodic case the
FMM is scaled approximately as $O(N)$ (with some small $N\log N$ addition
due to data structures), the present algorithm is scaled sublinearly at
small $N$, while it approaches the same scaling as the regular FMM.
Qualitatively this can be explained by complexity of the FMM as it used in
the present algorithm for periodization (see Eq. (\ref{o3}), where for
simplicity two last terms due to generation and evaluation of expansions are
neglected). So the ratio of the FMM time in the present algorithm and the
FMM for non-periodic problem can be estimated as 
\begin{equation}
\frac{T_{FMM}^{(per)}}{T_{FMM}^{(non)}}\sim \sqrt{\frac{A_{gf}}{B_{gf}}%
\left( 1+\frac{2L\left( \lambda \right) }{N}\right) }.  \label{lt1}
\end{equation}
\begin{figure}[tbh]
\begin{center}
\includegraphics[width=4.0in]
{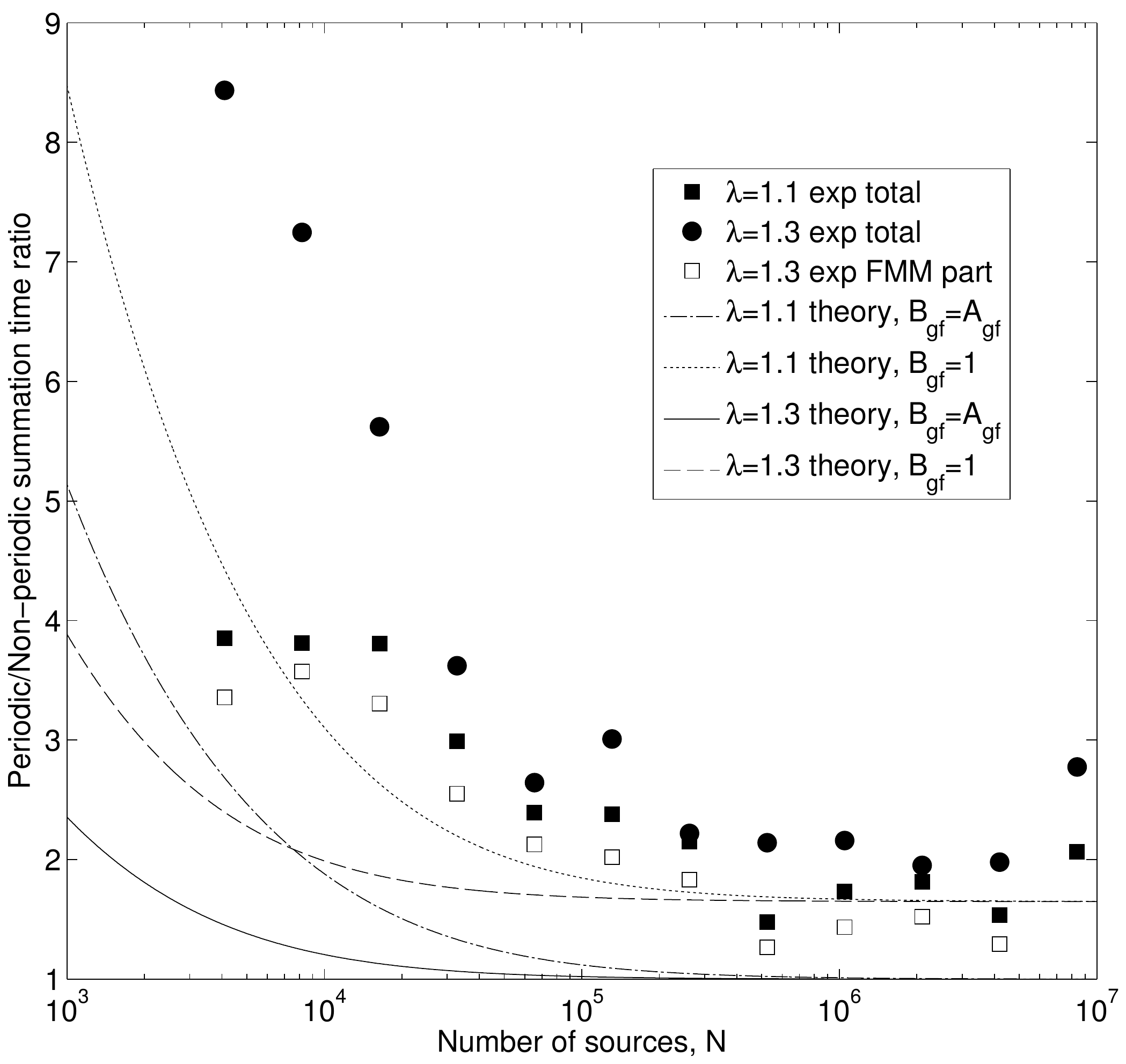}
\end{center}
\caption{The ratio of the solution times of the periodic and non-periodic
problems using the FMM for the cases shown in Fig. \ref{Performance}. Experimental results for the total time for the two cases, and the time taken by the FMM alone for the case $\lambda=1.3$ are shown. Also shown as continuous lines are theoretical estimates via Eq. (\ref{lt1}) for the same values, and different geometric factors.}
\label{TimeRatio}
\end{figure}

This ratio for different $\lambda $ along with the experimentally measured
time ratio of the ``get'' part of the algorithm and its FMM portion to the
FMM for non-periodic problem is plotted in Fig. \ref{TimeRatio}. For each $%
\lambda $ we presented theoretical ratios (\ref{lt1}) computed for limiting
cases $B_{gf}=1$ and $B_{gf}=A_{gf}$ \ (for cubic boxes $A_{gf}\approx 2.72$%
, see Eq. (\ref{o2.1})). It is seen that qualitatively this explains the
observed results, and at relatively small $N$ the wall clock time of the
present algorithm requires is several times larger than the time of
non-periodic FMM. At larger times the ratio should come to an asymptotic
limit depending on $\lambda $. Despite Eq. (\ref{o3}) predicts that $\lambda
=1.1$ limit should be somehow smaller than $\lambda =1.3$ limit, we found
that in the range of experiments they are approximately the same (time ratio
about 2, while can be 1.5 or so). The reason why the experimental data
deviate from Eq. (\ref{lt1}) is that there are some $O(N)$, $O(PN),$ and $%
O(P^{2})$ overheads in the periodization algorithm. The contribution of
these overheads is seen from the difference of the FMM part of the get
algorithm and the total cost. We also profiled the FMM, measured performance
constants, and found that at large $N\gtrsim 10^{6}$ cost approximation (\ref
{o3}) at $B_{gf}=A_{gf}$  is good enough and the FMM part approaches the
asymptotic constant, which for cubic domains can be, say, 1.2 ($p_{FMM}=16$%
). 

As the present algorithm can easily treat periodicity for non-cubic domains
we conducted accuracy and perfromance tests for periodic boxes of different
aspect ratios. The accuracy tests were performed by comparison of the
obtained solutions with the results of the Ewald summation in the same way
as described above in the range of $N$ from 1000 to 50000. It was found that
for fixed $\lambda $ and respective $p\left( \lambda \right) $ the box
aspect ratio practically does not affect the accuracy. At large aspect
ratios like 5:1:1 we found that conditioning of the system matrix $\mathbf{A}
$ (see Eq. (\ref{10})) can be poor, but this does not affect the accuracy of
the final results and a good algorithm for computing of pseudoinverse can
handle that. Some results of performance tests for are shown in Table 5.

Ideally, the time for non-periodic FMM at large $l_{\max }$ should not
depend on the box aspect ratio. However, some variation in time is seen,
which can be refered to non-perfect (discrete) optimization. For large box
aspect ratios one can get $N_{b}$ orders of magnitude larger than $N$ (e.g.
two orders for the case 10:1:1). For such cases the cost of computation of
the mean, which depends on $N_{b}$ becomes substantial. If these optional
computations are not needed, then performance improves, but anyway, solution
of periodic problem can be several times more expensive than solution of
non-periodic problem. For the cases presented in the table the ratio of
these times is well correlated with factor $A_{gf}^{1/2}$. This is
consistent with the principal term of the FMM cost (\ref{o3}) at $B_{gf}=1.$

{\small 
\begin{table}[tbh]
\caption{Performance for different box aspect ratios at $N=10^{5}$, $\lambda =1.3$, $\ p=34,$ $p_{FMM}=16.$ }
\label{Table5}
\begin{center}
{\small 
\begin{tabular}{|l|l|l|l|l|l|l|l|} \hline 
$d_{1}:d_{2}:d_{3}$ & 1:1:1 & 2:1:1 & 5:1:1 & 10:1:1 & 2:2:1 & 5:5:1 & 10:2:1
\\  \hline
$A_{gf}$ & 2.72 & 3.85 & 14.7 & 53.9 & 3.53 & 7.63 & 28.2 \\  \hline
$N_{b}$ & 597,625 & 845,188 & 3,228,941 & 11,849,769 & 776,783 & 1,675,371 & 
6,192,751 \\  \hline
$T^{(get)},$ s & 9.08 & 9.86 & 20.8 & 44.5 & 9.92 & 13.9 & 29.8 \\  \hline
$T_{(no\_mean)}^{(get)},$ s & 8.22 & 8.77 & 16.6 & 29.0 & 8.92 & 11.7 & 21.8
\\  \hline
$T_{(no\_mean)}^{(get)}/A_{gf}^{1/2},$ s & 4.98 & 4.47 & 4.33 & 3.94 & 4.74
& 4.24 & 4.11 \\  \hline
$T_{FMM}^{(non)},$ s & 3.76 & 4.71 & 4.03 & 3.40 & 3.45 & 5.65 & 4.79 \\ \hline
\end{tabular}
}
\end{center}
\end{table}
}

Finally we note that it is not an easy task to compare the absolute
performance of the present methods with the smooth particle mesh Ewald
(SPME) and other algorithms for periodic summation due to a difference in
implementation, accuracy, what is actually computed (potential and
gradient), hardware, etc. However, some comparison with that approaches can
be done using published data \cite{Kia:JCP2008} comparing performance of the
SPME and FMM-type PWA implementation for clusters, for relatively small size
problems ($N=10^{4}$ and $N=10^{5}$). The absolute figures indicate that the
wall clock time of the present algorithm for these sizes is of the same
order as the reported times for those methods, while we are able to perform
computations with larger problem sizes on relatively modest hardware. 

\subsection{Using graphics processing units (GPUs) for summation}

GPUs are often used to accelerate molecular dynamics simulations where
periodization may be required. We implemented the ``get'' part of the
algorithm completely on the GPU. Using ``brute-force'' summation on the GPU
and on a multicore CPU as the FSA, we did the same performance tests as for
the FMM, see Fig. \ref{Performance}. The ''set'' part can be executed on CPU
and transferred to the GPU. All parts of the algorithm are highly
parallelizable and so while the scaling of the algorithm is quadratic, it is
up to two orders of magnitude faster than the multicore CPU version.

Our tests reveal that at large truncation numbers $p$ (typically $p\gtrsim
30 $) single precision GPU computations cannot be used for spherical basis
function evaluation due to loss of precision in least-squares solution.
Double-precision computations provide accurate results with $L_{2}$-norm
relative errors of the order $10^{-13}-10^{-15}$ in potential and gradient
computations compared to the double precision CPU computations. Accordingly
Fig. \ref{Performance}, only shows double precision times. High performance
implementation of the brute-force summation is described in \cite
{Gumerov:JCP2008} and was used in the present tests with single and double
precision, and run on a single NVIDIA Tesla C2050 card. The CPU wall clock
times used for comparisons were measured for algorithm parallelized for 4
core PC described before.

Fig. \ref{Performance} shows the results of tests. It is seen that despite
the asymptotic quadratic scaling of the algorithm the GPU implementation can
be faster or comparable in speed with the FMM running on CPU at $p_{FMM}=16$
for problems of size $N\lesssim 10^{5}$. The ratio of times for solution of
periodic and non-periodic problems for brute force computations at large $N$
tends to theoretical limit $T_{brute}^{(per)}/T_{brute}^{(non)}=\pi \sqrt{%
\frac{3}{4}}\lambda ^{3}$, which for $\lambda =1.3$ used for illustrations
is about 6 times. Of course, this ratio can be reduced by decreasing $%
\lambda $. However, in the range $N\lesssim 10^{5}$ reduction of $\lambda $
below 1.1 does not benefit the overall performance due to increase of the
size of the expansion and reduction of the performance of evaluation of the
far field on the GPU, where the local memory is substantially smaller than
the CPU cache. One of the advantages of brute force double precision GPU
computing for problems of relatively small size ($N\lesssim 10^{5}$) is that
besides the roundoff errors the accuracy is controlled only by parameters $p$
and $\lambda $, while for the FSA=FMM the error is controlled also by $%
p_{FMM}$. For efficient FMM implementations on GPU this number is usually
small \cite{Gumerov:JCP2008} while the efficiency of the FMM on GPU for high
accuracy simulations is a subject for a separate study.

\section{Discussion}

By proposing a method for computation of periodic sums we do not mind to
compete with or diminish the role of the existing methods, such as based on
Ewald summation or the FMM for periodic functions for well-studied kernels.
We realize that a variety of problems with periodic boundary conditions and
practical concerns of researchers may require some solutions different from
the existing approaches, as any of them, including the present one has some
advantages and disadvantages. Such problems are not limited with Laplacian
kernels. However, development of specialized approaches which usually are
more efficient than generic approaches (kernel-independent FMM is an
example), also have some cost and availablity of a method which turns an
available FSA to its periodic version without any intervention to the basic
FSA algorith, in our opinion, is useful.

 The  ``cons'' and ``pros'' of the FFT vs FMM
were discussed in the literature, including performance and scaling at large 
$N$ (e.g. \cite{Yokota:CPC2013}) and application to solution of molecular
dynamics problems (e.g. \cite{Skeel:ICNAAM2010}). Particularly, some
difficulties for the use of the FMM for the MD simulations were reported as
a long term energy drift was observed in some computations using four or six
term expansion in the FMM \cite{Bishop:JCC1997} (perhaps, $p_{FMM}=4$ or $6$%
, which provides $p_{FMM}^{2} =16$ and 36 terms of expansion, respectively)$.
$ Of course, the number of terms should be evaluated carefully and the FMM
and other errors should be consistent with the integration schemes used
(particularly, the FMM errors in gradients are higher than in the potential 
\cite{Gumerov:JCP2013}). In this context, we can mention that in the present
algorithm one can enforce some conditions (such as setting the mean of the
potential to zero, similarly to the Ewald scheme), which is normally is not
controlled in the FMM-based algorithms. It can be more stable or not in
terms of the energy drift if applied to the MD simulations, but this
requires an additional study, which, certainly, goes beyond the scope of the
present work, where not only the MD applications are envisioned. Another advantage of not
requiring the FFT may arise for very large problems, which have to be implemented on distributed architectures, where the communication costs of the FMM are smaller than the FFT.

As far as comparing the present approach with an FMM algorithm implemented using periodic kernel functions is considered, we perform this comparison on three
sizes of problems: with small $N\ $(say, $N\lesssim 10^{3}$),
moderate $N$ (say, $10^{4}\lesssim N\lesssim 10^{5}$) and large $N$ (say, $%
N\gtrsim 10^{6}$). The first class may not require the FMM at
all, as direct sums may be done faster and be more accurate, and the
present algorithm can work efficiently and accurately (choosing proper $p$) with direct sums. The second and the third types of the problems require a fast summation algorithm such as the FMM. In this
case some authors claim that the overheads for solution of periodic problems
are negligibly small compared to solution of the non-periodic problem. For
example, Ref.  \cite{Amisaki:JCC2000} claims that periodization of
 requires only 0.1\% of the total
computational time. This statement {\em does not mean} that 
the periodic version of the sum is 1.001 times slower than the solution of the
non-periodic sum, as was remarked on an early draft of this paper. Indeed, in \cite{Amisaki:JCC2000} (page 1084 and Table V) it is stated ``{\em The time spent on dealing with periodicity ranged from about 0.3 to 2.6 times of that of the free boundary computations, depending on the values of $p$ and $L$ and the type of the FMM.}'' (Here $L=l_{\max}$ in our notation). This shows that actual measured time ratios, such as reported in Fig. \ref{TimeRatio}, were in the range 1.3 to 3.6. 

The overhead comes from the fact that the periodic FMM is the FMM executed
in the extended domain, which is not the original cell $\Omega _{0}$, but in 
$\Omega _{0}$ with its 26 neighbors, which is $\Omega _{b}$ in our notation,
see Fig. \ref{Fig1}. The difference between the present and periodic FMMs is
that, first, the periodic FMM saves on generation of multipole expansions
and multipole-to-multipole translations for the replicated domains, which is
not implemented in the present algorithm, and, second, that for the present
FMM some relatively small amount of receivers (``check points'') are added.
Discussions of the FMM costs and simplifications in Appendix A and around
Eqs (\ref{o3}) and (\ref{lt1}) show that asymptotically as $N\rightarrow
\infty $ the second modification brings a negligible overhead, while the
overhead due to the first modification is non-zero, and in practice can be
several percents for cubic $\Omega _{0}$ to tens of percents for
non-cubic $\Omega _{0}$ (due to the geometric factor $A_{gf}$ ), which
is the price of use of the FMM as a black box FSA. 

At $N\rightarrow \infty $ the costs of the periodic and
non-periodic FMMs are the same. The difference in costs of periodic and non-periodic FMMs
comes from two factors -- the difference in the multipole-to-local
translations and direct summations for the boxes near the boundaries of $%
\Omega _{0}$, due to the difference in the neighborhoods. We carefully
counted these numbers accounting for the boundary effects, and found that
for levels $l_{\max }=3$ and $4$ (midsize problems) used in \cite
{Amisaki:JCC2000}, the ratios of numbers of translations in periodic and
non-periodic FMMs is 1.96 and 1.38, while the ratios of the numbers of
direct summations is 1.30 and 1.14, respectively. At levels $l_{\max }=5$
and $6$ one can expect time ratios in a range 1.05-1.20.

One of the features of the present algorithm, which may be of practical
interest, is the capability to obtain periodic solutions for non-cubic domains
easily, which can be an issue for some other algorithms. These were illustrated in Table \ref{Table5}. Performance  for moderate
box aspect ratio is comparable to the cubic case. For high aspect ratios, the algorithm works, but some issues, which could be the subject of future study must be noted. There is a substantial increase in the amount of sources added because of the enclosure of the whole domain in a sphere. Because of this, both the memory needed  and  the
computational time are relatively large, while still scaled as expected. Possible ways to address this would be to devise more check point sets located not on a sphere, but on surface that are more conformal to the boundaries, and also a reduction of the
size of domain $\Omega _{0}$. The latter can be achieved in a few ways
including distribution of RBF centers not too far from the boundaries of $%
\Omega _{0}$ or to use basis functions, which are better suited for boxes of high
aspect ratios than the spherical basis functions.

\section{Conclusion}

We have presented a kernel-independent method for the periodization of
finite sums. The technique was presented in a general setting, and then
applied to the particular case of the Laplacian kernel using different
expansion bases. Tests showed that the method can be tuned to compute
periodic sums with arbitrary prescribed accuracy. In the case of use of the
FMM as the fast summation algorithm the complexity of the method at large $N$
is the same as the FMM. The computational time for large $N$ (in tests up to 
$N=2^{23}$) is about twice that for the finite box sum using the FMM.
Similar results are seen for GPU based FSA, though here the scaling is
quadratic, and the largest problem size that can be reasonably treated is
about $10^{5}$. The ease of implementation of the periodization method, its
performance, and capability to ``retrofit'' any available black box FSA
without any modification makes it practical. This method may also be
valuable on distributed architectures on which communication costs of an
algorithm are as important as computational complexity. FMM-based approaches
are known to be much more communication efficient than FFT-based approaches
for solution of the same large problems on distributed architectures.
Additional speedups of the method can be achieved by specialization of the
FMM -- these were specifically avoided in this paper to demonstrate the
ability to use a blackbox sum algorithm, and should be investigated if the
method is to be used in a ``production'' environment. Application to other
kernels should be straightforward, though details will have to be worked out
for them.

\appendix

\section{Complexity of the FMM}

There are many papers evaluating complexity of the FMM in basic settings
(mostly for the cases when the sources and receivers are the same and
uniformly occupy all computational domain, which can be thougth as a unit
cubic box, e.g. \cite{Gumerov:TR2005}). Here, we consider the FMM for
Laplace-like kernels and modify the problem by assuming that $N_{b}$
sources are located in the domain $\Omega _{b}$ of volume $V_{b}$ and $M$
receivers are distributed inside the domain $\Omega _{r}\subset \Omega _{b}$ of
volume $V_{r}.$ Let now $U$ be a minimal cube which includes $\Omega _{b}$
and presents the zero level of the octree. The cube is partitioned via the
octree down to level $l_{\max }$, in which case the maximum number of
sources in a box is $s$. Assuming more or less uniform source point
distribution in $\Omega _{b}$ and that any box in $\Omega _{r}$ contains at
least one receiver point, we can determine the number of source and receiver
boxes at this level as $B_{s}=N_{b}/s$ and $B_{r}=B_{s}V_{r}/V_{s}$,
respectively. The latter estimate comes from the observation that for large
enough $l_{\max }$ the ratio of the numbers of source and receiver boxes is
approximately the same as the ratio of volumes of the domains occupied by
the sources and receivers. In a standard FMM in three dimensions the number
of multipole-to-local translations to each receiver box located far enough
from the boundary of $\Omega _{b}$ is $K=189$ (this number can be reduced, e.g. as in \cite{Gumerov:JCP2008},
but anyway, $K\gg 1$). When counting the total number of translations $K$
should be increased to account for one local-to-local transaltion per
receiver box and 1 multipole-to-multipole translation per source box. It can be noted then that the amount of translations at the $l$th level of the 
octree increases geometrically with the level (8 times in three
dimensions), so the total number of translations, $N_{trans}$, will have
factor $8/7$ compared to the number of translations at level $l_{\max }$ and
we have 
\begin{equation}
N_{trans}\sim \frac{8}{7}\left( B_{s}+KB_{r}\right) =\frac{8N_{b}}{7s}\left(
1+K\frac{V_{r}}{V_{s}}\right) \approx \frac{8N_{b}}{7s}\frac{KV_{r}}{V_{s}}.
\label{F1}
\end{equation}
Here the latter approximate equality holds when $K\gg V_{s}/V_{r}$. The
number of direct summations per receiver is 27$s$, if we assume that the
neighborhood of the receiever box consists of 27 boxes, which yields the
total number of direct summations, $N_{dir}=27Ms.$ Hence, denoting $C_{trans}
$ the cost of a single translation and $C_{dir}$ the cost of a single direct
evaluation, we obtain the part of the total cost, which depends on $s$ in
the form 
\begin{equation}
C_{s}=\frac{8N_{b}}{7s}\frac{KV_{r}}{V_{s}}C_{trans}+27sMC_{dir}.  \label{F2}
\end{equation}
This function can be minimized with respect to  $s$, and the minimum cost denoted as, $%
C_{s,\min }$, at $s=s_{opt}$, where from $dC_{s}/ds=0$, we have 
\begin{equation}
s_{opt}=\sqrt{\frac{8KC_{trans}}{189C_{dir}}\frac{N_{b}}{M}\frac{V_{r}}{V_{s}%
}},\quad C_{s,\min }=C_{td}\sqrt{MN_{b}\frac{V_{r}}{V_{s}}},\quad C_{td}=4%
\sqrt{\frac{54K}{7}C_{trans}C_{dir}}  \label{F3}
\end{equation}
To get the total cost of the FMM, we need to add here the cost of generation
of multipole expansions, $C_{gen}N_{b}$ and the cost of evaluation of local
expansions, $C_{ev}M$, where $C_{gen}$ and $C_{ev}$ are the costs of
generation and evaluation of single expansions, which does not depend on $s$%
, or the depth of hte octree $l_{\max }$. Hence, the total cost of the FMM
used in the present algorithm, which neglects the cost of generation of data
structure is 
\begin{equation}
C^{(FMM)}=C_{td}\sqrt{MN_{b}\frac{V_{r}}{V_{s}}}+C_{gen}N_{b}+C_{ev}M.
\label{F4}
\end{equation}
Here the last estimate comes from the fact that usually translations and
direct summations are the major contributors to the overall cost of the FMM
(e.g. see profiling of the algorithm \cite{Gumerov:JCP2008}). Also note the
dependence of the FMM cost on the number of terms in the multipole and local
expansions used in the FMM, $P_{FMM}$. We have $C_{gen}\sim C_{gen}=O\left(
P_{FMM}\right) $, $C_{trans}=O\left( P_{FMM}^{\alpha }\right) ,$ and $%
C_{td}=O\left( P_{FMM}^{\alpha /2}\right) $ where $\alpha $ is some number
(e.g. for $O\left( p^{3}\right) $ translation methods, $\alpha =3/2$).

\section{Box integrals of the basis functions}

Basis functions $R_{n}^{m}\left( \mathbf{y}\right) $ are homogeneous
polynomials of degree $n$ (sums of monomials $x^{n_{1}}y^{n_{2}}z^{n_{3}},$ $%
n_{1}+n_{2}+n_{3}=1$). We can compute the integrals as 
\begin{eqnarray}
R_{t}^{(0)} &=&\frac{1}{V_{0}}\int_{\Omega _{0}}R_{t}\left( \mathbf{y}%
\right) dV\left( \mathbf{y}\right) =\frac{1}{d_{1}d_{2}d_{3}}%
\int_{-d_{1}/2}^{d_{1}/2}\int_{-d_{2}/2}^{d_{2}/2}%
\int_{-d_{3}/2}^{d_{3}/2}R_{t}\left( x,y,z\right) dxdydz  \label{A1} \\
&=&\frac{1}{8}\sum_{i=1}^{N_{q}}\sum_{j=1}^{N_{q}}%
\sum_{k=1}^{N_{q}}w_{i}w_{j}w_{k}R_{t}\left( \frac{1}{2}d_{1}x_{i},\frac{1}{2%
}d_{2}x_{j},\frac{1}{2}d_{3}x_{k}\right) ,  \notag
\end{eqnarray}
where $w_{i}$ and $x_{i}$ are the standard weights and abscissas of the
Gauss quadrature of order $N_{q}$ \cite{Abramowitz:Book1964}. Since this
integration is exact for polynomials of degree $n<2N_{q}$ and the maximum
degree of the polynomials in the sum is $p-1,$ the choice 
\begin{equation}
N_{q}=\left[ \frac{p-1}{2}\right] +1,\quad p=P^{1/2},  \label{A2}
\end{equation}
provides an exact result. For evaluation of all $P$ required coefficients $%
\widehat{R}_{n}^{m(0)}$ the computational cost of this procedure is $O\left(
P^{5/2}\right).$ Note that faster methods may be proposed for computation of
this step. However, this is not crucial, as this integration is performed in
the ``set'' part of the algorithm, which overall cost is $O\left(
P^{3}\right)$, and this cost is amortized over the rest of the algorithm.

\section{Mean computations}

To compute the integral (\ref{pfx4}) for the kernel (\ref{2}), we first
apply the Gauss divergence theorem to reduce the volume integral to a
surface integral: 
\begin{equation}
\Phi _{0}\left( \mathbf{x}\right) =\frac{1}{V_{0}}\int_{\Omega _{0}}\frac{%
dV\left( \mathbf{y}\right) }{\left| \mathbf{y}-\mathbf{x}\right| }=\frac{1}{%
2V_{0}}\int_{\Omega _{0}}\nabla _{\mathbf{y}}\cdot \left( \frac{\mathbf{y}-%
\mathbf{x}}{\left| \mathbf{y}-\mathbf{x}\right| }\right) dV\left( \mathbf{y}%
\right) =\frac{1}{2V_{0}}\int_{\partial \Omega _{0}}\frac{\mathbf{n}\cdot
\left( \mathbf{y}-\mathbf{x}\right) }{\left| \mathbf{y}-\mathbf{x}\right| }%
dS\left( \mathbf{y}\right) ,  \label{B1}
\end{equation}
where $\mathbf{n}$ is the outward normal to $\Omega _{0}.$ This result is
valid for an arbitrary point $\mathbf{x}$ including when $\mathbf{x}$ is
located in $\Omega _{0}$ or on its boundary $\partial \Omega _{0}$. This can
be checked by consideration of $\epsilon $-vicinities of singularities,
which are integrable. The surface integral can be decomposed into integrals
over the box faces, $S_{k}$, $k=1,...,6.$%
\begin{eqnarray}
\Phi _{0}\left( \mathbf{x}\right) &=&\frac{1}{2V_{0}}\sum_{k=1}^{6}%
\int_{S_{k}}\frac{\mathbf{n}_{k}\cdot \left( \mathbf{y}-\mathbf{x}\right) }{%
\left| \mathbf{y}-\mathbf{x}\right| }dS\left( \mathbf{y}\right) =-\frac{1}{%
2V_{0}}\sum_{k=1}^{6}\left( \mathbf{n}_{k}\cdot \mathbf{x}_{k}\right)
L_{k}\left( \mathbf{x}_{k}\right) ,\quad  \label{B2} \\
L_{k}\left( \mathbf{x}_{k}\right) &=&\int_{S_{k}}\frac{dS\left( \mathbf{y}%
_{k}\right) }{\left| \mathbf{y}_{k}-\mathbf{x}_{k}\right| },\quad \mathbf{y}%
_{k}=\mathbf{y-y}_{k0},\quad \mathbf{x}_{k}=\mathbf{x-y}_{k0},  \notag
\end{eqnarray}
where $\mathbf{n}_{k}$ and $\mathbf{y}_{k0}$ are the normal and the center
of the $k$th face, while $\mathbf{y}_{k}$ and $\mathbf{x}_{k}$ are
coordinates in the reference frame with the origin at the $k$th face center.
The surface integral then can be reduced to the contour integral using, e.g.
the Gauss divergence theorem in the plane of a particular face. Indeed,
consider function 
\begin{eqnarray}
\mathbf{F}_{k}\left( \mathbf{r}_{k}\right) &=&\mathbf{r}_{k}f_{k}\left(
r_{k};h_{k}\right) ,\quad f_{k}\left( r_{k};h_{k}\right) =\frac{\rho
_{k}-h_{k}}{r_{k}^{2}},\quad \rho _{k}=\sqrt{r_{k}^{2}+h_{k}^{2}}\left(
=\left| \mathbf{y}_{k}-\mathbf{x}_{k}\right| \right) ,\quad  \label{B3} \\
\quad \mathbf{r}_{k} &=&\mathbf{y}_{k}-\mathbf{x}_{k}^{\prime },\quad 
\mathbf{x}_{k}^{\prime }=\mathbf{x}_{k}-\mathbf{n}_{k}h_{k},\quad h_{k}=%
\mathbf{n}_{k}\cdot \mathbf{x}_{k}.  \notag
\end{eqnarray}
The 2D divergence of this function in the plane of the $k$th face is $1/\rho
_{k}$. So 
\begin{equation}
L_{k}\left( \mathbf{x}_{k}\right) =\int_{S_{k}}\widetilde{\nabla }_{\mathbf{r%
}_{k}}\cdot \mathbf{F}_{k}\left( \mathbf{r}_{k}\right) dS\left( \mathbf{r}%
_{k}\right) =\int_{C_{k}}\mathbf{n}_{k}^{\prime }\cdot \mathbf{F}_{k}\left( 
\mathbf{r}_{k}\right) dl\left( \mathbf{r}_{k}\right) ,  \label{B4}
\end{equation}
where $\mathbf{n}_{k}^{\prime }$ is the outer normal to the contour $%
C_{k}=\partial \Omega _{k}$. This integral can be decomposed into four
integrals over the face edges. So 
\begin{equation}
L_{k}\left( \mathbf{x}_{k}\right) =\sum_{j=1}^{4}I_{kj}\left( \mathbf{x}%
_{k}\right) ,\quad I_{kj}\left( \mathbf{x}_{k}\right) =\int_{C_{kj}}\mathbf{n%
}_{kj}^{\prime }\cdot \mathbf{r}_{k}\frac{\rho _{k}-h_{k}}{r_{k}^{2}}dl.
\label{B5}
\end{equation}
The latter integrals can be found analytically. Indeed, consider for the $j$%
th edge a local right hand oriented reference frame centered at its endpoint 
$\mathbf{y}_{kj0}$ from which integration starts, and unit basis vectors $%
\mathbf{i}_{kjx}^{\prime }$ directed along the integration path, $\mathbf{i}%
_{kjy}^{\prime }=\mathbf{n}_{k}$ and $\mathbf{i}_{kjz}^{\prime }=\mathbf{n}%
_{kj}^{\prime }=\mathbf{i}_{kjx}^{\prime }\times \mathbf{i}_{kjy}^{\prime }.$
Denoting coordinates of $\mathbf{x}$ in this reference frame as 
\begin{equation}
x_{kj}=\left( \mathbf{x}_{k}-\mathbf{y}_{kj0}\right) \cdot \mathbf{i}%
_{kjx}^{\prime },\quad y_{kj}=\left( \mathbf{x}_{k}-\mathbf{y}_{kj0}\right)
\cdot \mathbf{i}_{kjy}^{\prime },\quad z_{kj}=\left( \mathbf{x}_{k}-\mathbf{y%
}_{kj0}\right) \cdot \mathbf{i}_{kjz}^{\prime },  \label{B6}
\end{equation}
we obtain 
\begin{equation}
I_{kj}\left( \mathbf{x}_{k}\right)
=-z_{kj}\int_{-x_{kj}}^{l_{kj}-x_{kj}}f\left( r_{kj};\left| y_{kj}\right|
\right) dx=H\left( l_{kj}-x_{kj},\left| y_{kj}\right| ,z_{kj}\right)
-H\left( -x_{kj},\left| y_{kj}\right| ,z_{kj}\right) ,  \label{B7}
\end{equation}
where $r_{kj}^{2}=x^{2}+z_{kj}^{2}$, $l_{kj}$ is the length of edge $C_{kj}$%
, and $H\left( x,y,z\right) $ is the primitive, 
\begin{equation}
H\left( x,y,z\right) =-z\int f\left( r;\left| y\right| \right) dx,\quad
f\left( r;y\right) =\frac{\rho -\left| y\right| }{r^{2}},\quad
r^{2}=x^{2}+z^{2},\quad \rho =\sqrt{r^{2}+y^{2}},  \label{B8}
\end{equation}
which can be computed analytically as 
\begin{equation}
H\left( x,y,z\right) =\left| y\right| \left( \arctan \frac{x}{z}-\arctan 
\frac{\left| y\right| x}{z\rho }\right) -z\ln \left| \rho +x\right| +C\left(
y,z\right) .  \label{B9}
\end{equation}
The integration constant $C\left( y,z\right) $ can be selected arbitrarily
to eliminate possible singularities. Particularly for $y\neq 0,$ $z=0$ one
can set $H=0$. The above formulae are sufficient for numerical
implementation, which in the simplest form can program the primitive (\ref
{B9}) and implement the above decompositions. There exist some box
symmetries (e.g. all local coordinates are nothing but permuted and shifted
original Cartesian coordinates), which can be exploited to achieve better
performance.

\section{Ewald summation}

The Ewald summation method is based on decomposition of kernel (\ref{2}) 
\begin{eqnarray}
K\left( \mathbf{y}-\mathbf{x}\right) &=&K_{1}\left( \mathbf{y}-\mathbf{x};\;
\xi \right) +K_{2}\left( \mathbf{y}-\mathbf{x}; \; \xi \right) ,\quad
\label{e1} \\
K_{1}\left( \mathbf{y}-\mathbf{x}; \; \xi \right) &=&\frac{\text{erfc}\left(
\xi \left| \mathbf{y}-\mathbf{x}\right| \right) }{\left| \mathbf{y}-\mathbf{x%
}\right| },\quad \mathbf{y}\neq \mathbf{x;\quad }K_{1}\left( \mathbf{0}; \;
\xi \right) =-\frac{2\xi }{\sqrt{\pi }},  \notag \\
K_{2}\left( \mathbf{y}-\mathbf{x}; \; \xi \right) &=&\frac{\text{erf}\left(
\xi \left| \mathbf{y}-\mathbf{x}\right| \right) }{\left| \mathbf{y}-\mathbf{x%
}\right| },\quad \forall \mathbf{y},\mathbf{x}\in \mathbb{R}^{3},\quad
\left( K_{2}\left( \mathbf{0}; \; \xi \right) =\frac{2\xi }{\sqrt{\pi }}%
\right) ,  \notag
\end{eqnarray}
which is exact for any value of parameter $\xi $, since by definition of the
error function, erf$\left( x\right) $, and the complimentary error function,
erfc$\left( x\right) $, we have erf$\left( x\right) +$erfc$\left( x\right)
=1 $ and the value of $K_{1}\left( \mathbf{0;}\xi \right) $ is set due to by
definition $K\left( \mathbf{0}\right) =0.$ So for the total potential (\ref
{2}) we have 
\begin{eqnarray}
\phi \left( \mathbf{y}\right) &=& \phi _{1}\left( \mathbf{y}\right) +\phi
_{2}\left( \mathbf{y}\right) ,\quad \phi _{1}\left( \mathbf{y}\right) =\sum_{%
\mathbf{p}}\sum_{i=1}^{N}q_{i}K_{1}\left( \mathbf{y}-\mathbf{x}_{i}+\mathbf{p%
}; \; \xi \right) ,  \notag \\
\phi _{2}\left( \mathbf{y}\right) &=& \sum_{\mathbf{p}%
}\sum_{i=1}^{N}q_{i}K_{2}\left( \mathbf{y}-\mathbf{x}_{i}+\mathbf{p}; \; \xi
\right) .  \label{e2}
\end{eqnarray}
Both functions $\phi _{1}\left( \mathbf{y}\right) $ and $\phi _{2}\left( 
\mathbf{y}\right) $ are periodic.

Due to fast decay of erfc$\left( x\right) $ computation of $\phi _{1}\left( 
\mathbf{y}\right) $ for $\mathbf{y}\in \Omega _{0}$ can be done only using
the sources in some neighborhood of\ $\Omega _{0}$, namely in $\Omega
_{1}\supset \Omega _{0}$ such that the minimum distance, $a$, between the
points on the boundaries $\partial \Omega _{0}$ and $\partial \Omega _{1}$
is much larger than $1/\xi $. Hence, this can be computed directly by
evaluation of a finite sum with a controllable error as 
\begin{equation}
\phi _{1}\left( \mathbf{y}\right) =\sum_{\mathbf{x}_{j}\in \Omega _{1}\left(
\xi \right) }q_{j}K_{1}\left( \mathbf{y}-\mathbf{x}_{j};\;\xi \right)
+O\left( e^{-\xi ^{2}a^{2}}\right) .  \label{e3}
\end{equation}
For computation of $\phi _{2}\left( \mathbf{y}\right) $ one can notice that $%
K_{2}$ is a solution of the Poisson equation 
\begin{equation}
\nabla ^{2}K_{2}\left( \mathbf{y}-\mathbf{x};\;\xi \right) =-4\pi \delta
_{\xi }\left( \mathbf{y}-\mathbf{x}\right) ,\quad \delta _{\xi }\left( 
\mathbf{y}-\mathbf{x}\right) =\frac{\xi ^{3}}{\pi ^{3/2}}e^{-\xi ^{2}\left| 
\mathbf{y}-\mathbf{x}\right| ^{2}},  \label{e4}
\end{equation}
where $\delta _{\xi }\left( \mathbf{y}-\mathbf{x}\right) $ is a compactly
supported function, which turns to the Dirac delta-function as $\xi
\rightarrow \infty $. Periodic solution of the Poisson equation can be
obtained via the FFT. For this purpose, we grid the domain $\Omega _{0}$ and
select $\xi $ in a way that $\xi \ll 1/h$, and $\xi \gg 1/\max \left(
d_{1},d_{2},d_{3}\right) $ (an optimal setting can be found from analysis of
the error bounds), where $h$ is the minimum spatial step of the grid. This
enables sampling of $\delta _{\xi }\left( \mathbf{y}-\mathbf{x}\right) $ for
source $\mathbf{x=x}_{i}$ at several grid points around $\mathbf{x}_{i}$.
The number of these grid points determines the accuracy of the method (at
optimal settings), so we introduce additional parameter $N_{r}$, so $\delta
_{\xi }\left( \mathbf{y}\right) $ is sampled in a box $\left(
2N_{r}+1\right) \times \left( 2N_{r}+1\right) \times (2N_{r}+1)$. We also
take care about the points $\mathbf{x}_{i}$ located near the boundary of $%
\Omega _{0}$ by periodization (so we construct a periodic function $\delta
_{\xi }^{(\mathbf{p})}\left( \mathbf{y}-\mathbf{x}\right) $). Further, we
apply the forward 3D FFT to 
\begin{equation}
f_{2}\left( \mathbf{y}\right) =\nabla ^{2}\phi _{2}\left( \mathbf{y}\right)
=-4\pi \sum_{i=1}^{N}q_{i}\delta _{\xi }^{(\mathbf{p})}\left( \mathbf{y}-%
\mathbf{x}_{i}\right) ,  \label{e5}
\end{equation}
and zero the harmonic of the Fourier image $f_{2}^{\ast }\left( \mathbf{k}%
\right) $ corresponding to the wavenumber $k=0.$ The inverse 3D FFT of $\phi
_{2}^{\ast }\left( \mathbf{k}\right) =-k^{-2}f_{2}^{\ast }\left( \mathbf{k}%
\right) $, produces the required solution $\phi _{2}\left( \mathbf{y}\right) 
$ with zero mean at grid points. Note then that solution obtained in this
way has the following mean 
\begin{equation}
\phi _{mean}\left( \xi \right) =\left\langle \phi \left( \mathbf{y}\right)
\right\rangle _{\Omega _{0}}=\frac{1}{V_{0}}\sum_{\mathbf{x}_{j}\in \Omega
_{1}\left( \xi \right) }q_{j}\int_{\Omega _{0}}K_{1}\left( \mathbf{y}-%
\mathbf{x}_{j}\mathbf{;}\xi \right) dV\approx 0.  \label{e6}
\end{equation}
The zero mean here is due to the compact support of the kernel $K_{1}$ and
charge neutrality. This mean can be computed using decomposition $%
K_{1}\left( \mathbf{y}-\mathbf{x}_{j}\mathbf{;}\xi \right) =K\left( \mathbf{y%
}-\mathbf{x}_{j};\xi \right) -K_{2}\left( \mathbf{y}-\mathbf{x}_{j};\xi
\right) $, where the integral with the first kernel can be computed
analytically (see Appendix C), while the integral with the second kernel is
regular and can be computed using, say, the trapezoidal rule (in the
FFT-based method the space is gridded). To avoid interpolation errors, in
the numerical tests where we compared our method for accuracy with the Ewald
summation method, we used only cases when the source and evaluation points
are located at the grid nodes.

\end{document}